\documentclass[pdflatex,sn-nature]{sn-jnl}

\usepackage{graphicx}%
\usepackage{multirow}%
\usepackage{amsmath,amssymb,amsfonts}%
\usepackage{amsthm}%
\usepackage{mathrsfs}%
\usepackage[title]{appendix}%
\usepackage{xcolor}%
\usepackage{textcomp}%
\usepackage{manyfoot}%
\usepackage{booktabs}%
\usepackage{algorithm}%
\usepackage{algorithmicx}%
\usepackage{algpseudocode}%
\usepackage{listings}%

\newcommand\apj{Astrophys. J.} 
\newcommand\apjl{Astrophys. J. Lett.} 
\newcommand\apjs{Astrophys. J. Suppl.} 
\newcommand\mnras{Mon. Not. R. Astron. Soc.} 
\newcommand\aap{Astron. \& Astrophys.} 
\newcommand\pasj{Publ. Astron. Soc. Jpn}

\newcommand\nat{Nature.} 
\usepackage{caption}

\theoremstyle{thmstyleone}%
\theoremstyle{thmstyletwo}%

\theoremstyle{thmstylethree}%

\begin{document}

\title[Article Title]{Vigorous turbulence driven by quasar-mode feedback in a cluster core}


\author*[1,2,3,4]{\fnm{Satoshi} \sur{Yamada}}\email{satoshi.yamada@astr.tohoku.ac.jp}

\author*[5,6]{\fnm{Shutaro} \sur{Ueda}}\email{shutaro@se.kanazawa-u.ac.jp}

\author*[3]{\fnm{Hirofumi} \sur{Noda}}\email{hirofumi.noda@astr.tohoku.ac.jp}

\author*[7]{\fnm{Yutaka} \sur{Fujita}}\email{y-fujita@tmu.ac.jp}

\author[8]{\fnm{Misaki} \sur{Mizumoto}}

\author[9,10,11,12,13]{\fnm{Kentaro} \sur{Nagamine}}

\author[2,14,15]{\fnm{Claudio} \sur{Ricci}}

\author[16,17]{\fnm{Shoji} \sur{Ogawa}}

\author[18,4]{\fnm{Taiki} \sur{Kawamuro}}

\author[19]{\fnm{Shinya} \sur{Yamada}}

\author[20]{\fnm{Yuichi} \sur{Terashima}}

\author[21]{\fnm{Yoshihiro} \sur{Ueda}}

\affil[1]{\orgdiv{Frontier Research Institute for Interdisciplinary Sciences}, \orgname{Tohoku University}, \orgaddress{\city{Sendai}, \postcode{980-8578}, \country{Japan}}}

\affil[2]{\orgdiv{Department of Astronomy}, \orgname{University of Geneva}, \orgaddress{\street{ch.d’Ecogia 16}, \city{Versoix}, \postcode{1290}, \country{Switzerland}}}

\affil[3]{\orgdiv{Astronomical Institute, Graduate School of Science}, \orgname{Tohoku University}, \orgaddress{\city{Sendai}, \postcode{980-8578}, \country{Japan}}}

\affil[4]{\orgname{RIKEN Cluster for Pioneering Research}, \orgaddress{\street{2-1 Hirosawa}, \city{Wako}, \postcode{351-0198}, \country{Japan}}}

\affil[5]{\orgdiv{Faculty of Mathematics and Physics, Institute of Science and Engineering}, \orgname{Kanazawa University}, \orgaddress{\street{Kakuma}, \city{Kanazawa, Ishikawa}, \postcode{920-1192}, \country{Japan}}}

\affil[6]{\orgdiv{Advanced Research Center for Space Science and Technology, College of Science and Engineering}, \orgname{Kanazawa University}, \orgaddress{\street{Kakuma}, \city{Kanazawa, Ishikawa}, \postcode{920-1192}, \country{Japan}}}

\affil[7]{\orgdiv{Department of Physics, Graduate School of Science}, \orgname{Tokyo Metropolitan University}, \orgaddress{\street{1-1 Minami-Osawa}, \city{Hachioji-shi, Tokyo}, \postcode{192-0397}, \country{Japan}}}

\affil[8]{\orgdiv{Science Education Research Unit}, \orgname{University of Teacher Education Fukuoka}, \orgaddress{\street{Munakata}, \city{Fukuoka}, \postcode{811-4192}, \country{Japan}}}

\affil[9]{\orgdiv{Theoretical Astrophysics, Department of Earth and Space Science}, \orgname{The University of Osaka}, \orgaddress{\street{1-1 Machikaneyama}, \city{Toyonaka, Osaka}, \postcode{560-0043}, \country{Japan}}}

\affil[10]{\orgdiv{Theoretical Joint Research, Forefront Research Center}, \orgname{The University of Osaka}, \orgaddress{\street{1-1 Machikaneyama}, \city{Toyonaka, Osaka}, \postcode{560-0043}, \country{Japan}}}

\affil[11]{\orgdiv{Kavli IPMU (WPI), UTIAS}, \orgname{The University of Tokyo}, \orgaddress{\city{Kashiwa, Chiba}, \postcode{277-8583}, \country{Japan}}}

\affil[12]{\orgdiv{Department of Physics and Astronomy}, \orgname{University of Nevada}, \orgaddress{\street{Las Vegas, 4505 S. Maryland Pkwy}, \city{Las Vegas}, \postcode{NV 89154-4002}, \country{USA}}}

\affil[13]{\orgdiv{Nevada Center for Astrophysics}, \orgname{University of Nevada}, \orgaddress{\street{Las Vegas, 4505 S. Maryland Pkwy}, \city{Las Vegas}, \postcode{NV 89154-4002}, \country{USA}}}

\affil[14]{\orgdiv{Instituto de Estudios Astrofísicos, Facultad de Ingeniería y Ciencias}, \orgname{Universidad Diego Portales}, \orgaddress{\street{Av. Ejército Libertador 441}, \city{Santiago}, \country{Chile}}}

\affil[15]{\orgdiv{Kavli Institute for Astronomy and Astrophysics}, \orgname{Peking University}, \orgaddress{\street{Beijing 100871}, \country{China}}}

\affil[16]{\orgdiv{Faculty of Science and Technology}, \orgname{Tokyo University of Science}, \orgaddress{\street{2641 
Yamazaki}, \city{Noda}, \postcode{278-8510}, \state{Chiba}, \country{Japan}}}

\affil[17]{\orgdiv{Institute of Space and Astronautical Science (ISAS)}, \orgname{Japan Aerospace Exploration Agency (JAXA)}, \orgaddress{\street{Sagamihara}, \city{Kanagawa}, \postcode{252-5210}, \country{Japan}}}

\affil[18]{\orgdiv{Department of Earth and Space Science}, \orgname{The University of Osaka}, \orgaddress{\street{1-1 Machikaneyama}, \city{Toyonaka}, \postcode{560-0043}, \state{Osaka}, \country{Japan}}}

\affil[19]{\orgdiv{Department of Physics}, \orgname{Rikkyo University}, \orgaddress{\street{3-34-1 Nishi Ikebukuro, Toshima-ku}, \city{Tokyo}, \postcode{171-8501}, \country{Japan}}}

\affil[20]{\orgdiv{Department of Physics, Faculty of Science}, \orgname{Ehime University}, \orgaddress{\street{2-5 Bunkyo-cho, Matsuyama}, \city{Ehime}, \postcode{790-8577}, \country{Japan}}}

\affil[21]{\orgdiv{Department of Astronomy}, \orgname{Kyoto University}, \orgaddress{\street{Kitashirakawa-Oiwake-cho, Sakyo-ku}, \city{Kyoto}, \postcode{606-8502}, \country{Japan}}}


\abstract{
Quasars are among the most luminous objects. 
They are powered by accretion onto supermassive black holes. 
They are thought to impact cosmological evolution primarily through energetic winds, known as quasar-mode feedback, yet the efficiency and spatial extent of this process remain poorly constrained. 
Here we present X-Ray Imaging and Spectroscopy Mission (XRISM) observations of H1821+643---the nearest galaxy cluster with a central quasar (redshift $z = 0.297$)---which was a rare opportunity to directly probe quasar-mode feedback in the intracluster medium. 
High-resolution spectroscopy reveals exceptionally broadened Fe\,\textsc{xxv} emission lines from the intracluster medium, with a velocity dispersion of approximately 300~km~s$^{-1}$, far exceeding values observed in nearby cluster cores. 
These lines originate predominantly at radii of 20--100~kpc from the centre.
Assuming that turbulence from a quasar-driven shock led to the broadening of the lines, the energy injected by the quasar beyond galactic scales ($\gtrsim$20~kpc) is estimated to be $\gtrsim$1--10\% of its radiative energy. 
Notably, this feedback efficiency exceeds previous multiwavelength estimates by orders of magnitude ($\lesssim$0.01\%) and reaches the levels required by the latest cosmological hydrodynamical simulations.
This finding of vigorous turbulence indicates that quasar-mode feedback plays a central role in regulating galaxy and cluster evolution at high redshift.
}

\maketitle



The X-Ray Imaging and Spectroscopy Mission (XRISM) \cite{Tashiro2025} is the successor to the Hitomi X-ray observatory \cite{Takahashi2014}, which ceased operations prematurely in 2016.
The Resolve instrument onboard XRISM is a cryogenically cooled, non-dispersive X-ray microcalorimeter that provides an energy resolution of 4.5~eV full-width at half-maximum, with a gain uncertainty of $<$0.3~eV at 5.9~keV, equivalent to a velocity resolution of better than $<$15~km~s$^{-1}$ (ref. \cite{Eckart2025}). 
The $6 \times 6$~px$^2$ array of Resolve covers a $3' \times 3'$ field of view with a half-power diameter of ${\sim}1.3'$. 
These capabilities enable direct measurements of the thermodynamic and dynamic properties of diffuse X-ray-emitting plasma, such as the intracluster medium (ICM) in galaxy clusters, just as Hitomi observed the Perseus cluster \cite{Hitomi2016,Hitomi2017}.

H1821+643 ($z = 0.29708$ \cite{Bahcall1992}), the nearest galaxy cluster hosting a central quasar \cite{Russell2010,Hlavacek-Larrondo2013}, is an ideal target for examining the impact of quasar activity on the surrounding ICM. 
Hitomi and recent XRISM observations have shown that turbulent motions in the ICM core are generally moderate, namely ${\lesssim}$160~km~s$^{-1}$ in nearby cluster cores without undergoing large mergers, even though they host active galactic nuclei (AGNs) launching relativistic jets \cite{Hitomi2016,XRISM25_Abell2029_I,XRISM25_Abell2029_II,Rose2025,XRISM26_Virgo,Fujita2025,XRISM25_Centaurus}.
H1821+643 hosts an extremely powerful AGN classified as a radio-quiet quasar. 
The AGN has a bolometric luminosity $L_{\rm QSO} \simeq (1$--$2) \times 10^{47}$~erg~s$^{-1}$ (refs. \cite{Russell2010,Gupta2024}), making it 1--5 orders of magnitude more luminous than other quasars (Supplementary Table\,\ref{Sup_Tab1}), while hosting only weak FR I-like twin radio jets \cite{Blundell2001}.
The black hole mass estimated by reverberation mapping is $M_{\rm BH} \simeq 2.6 \times 10^{9} M_{\odot}$ ($M_{\odot}$ is the solar mass) \cite{Shapovalova2016}.
Previous multiwavelength observations have revealed that quasars can drive galactic-scale energetic winds \cite{Fiore2017}, but there is almost no evidence for substantial energy injection beyond galactic scales ($\gtrsim$20~kpc) \cite{Ruan2015,Lacy2019,Venturi2023,Harrison2024}.
To assess the influence of the quasar-mode feedback on cluster-scale environments, we observed H1821+643 with XRISM between 4 and 10 September 2024, obtaining a net exposure of 283.7~ks.



\begin{figure}[t]
    \centering
    \includegraphics[width=\textwidth]{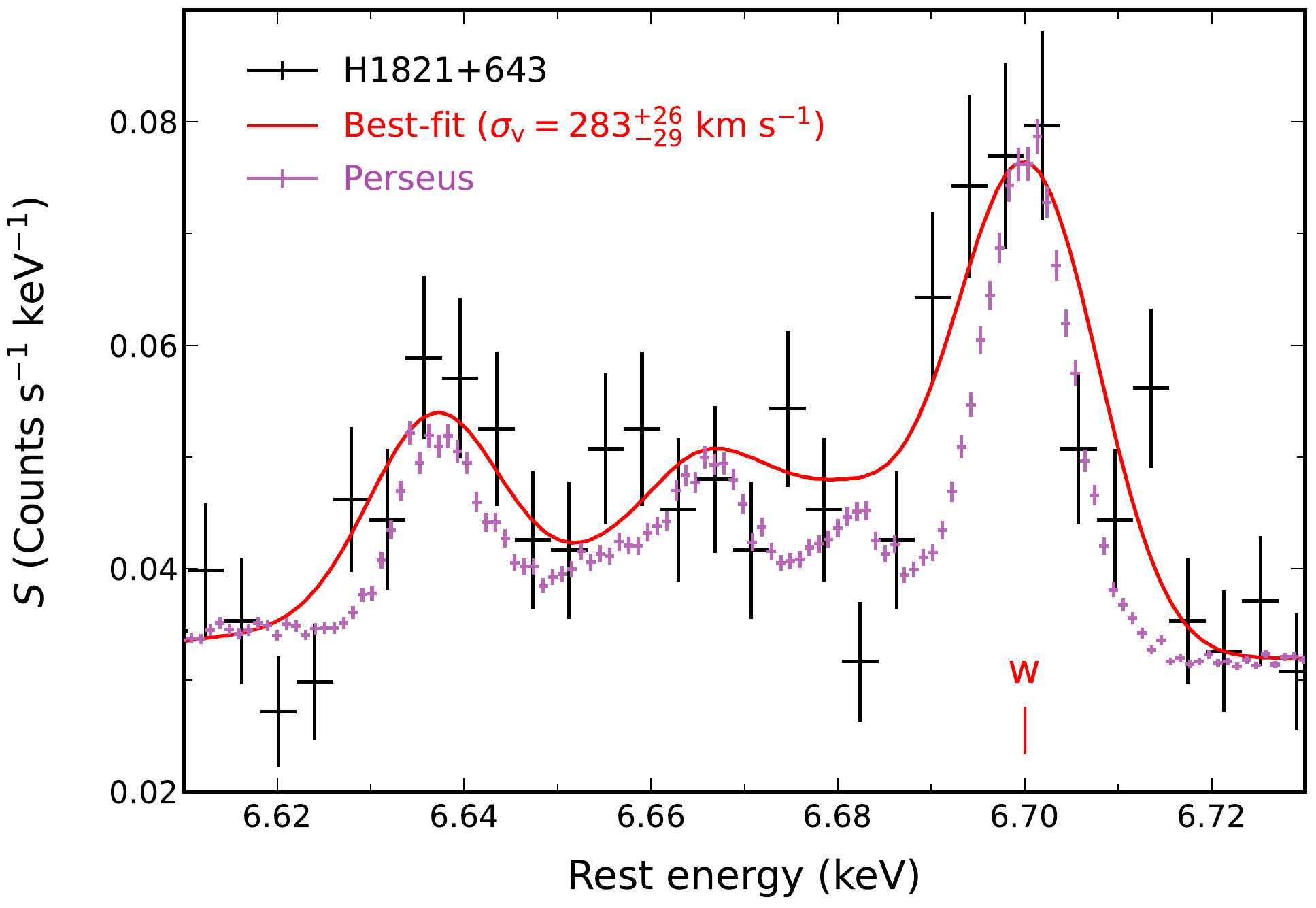}
    \vspace{0.03 cm}
    \caption{{\label{Fig1}\bf Fe\,\textsc{xxv} He$\alpha$ spectrum from the ICM in H1821+643.}
    XRISM/Resolve spectrum (black) and the best-fitting model (red) are shown in the rest frame of the source ($z_{\rm ICM} = 0.2973$). 
    For comparison, the Hitomi spectrum of the Perseus cluster core ($z_{\rm ICM} = 0.017284$) is shown in purple, normalized to match the continuum level and peak line intensity of H1821+643. 
    Uncertainties correspond to $1\sigma$. 
    The resonance (w) line (marked by a short red line) in H1821+643 has a broader profile than that of Perseus ($\sigma_{v} \approx 164$~km~s$^{-1}$), indicating substantially stronger ICM turbulence and velocity shear.
    }
\end{figure}

Figure\,\ref{Fig1} shows the redshift-corrected X-ray spectrum of the Fe\,\textsc{xxv} He$\alpha$ emission lines from the ICM in H1821+643 as observed with Resolve and extracted from the full field of view. 
For comparison, we overlaid the Hitomi spectrum of the Perseus cluster, which hosts a less luminous AGN \cite{Hitomi2016}. 
The linewidth is substantially broader than that in the Perseus cluster (Fig.\,\ref{Fig1}). 
We modelled the observed 4--7-keV spectrum (rest frame 5.2--9.1~keV), which includes Fe\,\textsc{xxv} He$\alpha$, He$\beta$, He$\gamma$ and Fe\,\textsc{xxvi} Ly$\alpha$ (Methods and Extended Data Fig.\,\ref{Ext_Fig1}).
The model consists of collisionally ionized plasma emission for the ICM and a power-law emission and a neutral Fe K$\alpha$ line for the quasar, yielding a line-of-sight velocity dispersion of $\sigma_{v} = 295^{+34}_{-28}$~km~s$^{-1}$. 
Next, we masked the Fe\,\textsc{xxv} He$\alpha$ lines in the spectral analysis and refitted the spectrum, thereby obtaining a consistent velocity dispersion of $\sim$300~km~s$^{-1}$ (Extended Data Fig.\,\ref{Ext_Fig2}). 
However, the observed flux of the resonance (w) line was smaller than predicted by the model, indicating potential resonance scattering \cite{Gilfanov1987,Hitomi2018_RS}. 
To account for the resonance scattering, we included a Gaussian absorption component for the w-line, yielding $\sigma_{v} = 283^{+26}_{-29}$~km~s$^{-1}$ and a temperature $kT = 6.2^{+0.3}_{-0.4}$~keV. 
In summary, even after accounting for systematic uncertainties, the observed velocity dispersion ($\sim$300~km~s$^{-1}$) is substantially higher than that observed in nearby cluster cores hosting less luminous AGNs observed with XRISM and Hitomi (${\lesssim}160$~km~s$^{-1}$; Methods), including the Perseus cluster (${\approx}164$~km~s$^{-1}$; ref. \cite{Hitomi2016} and Fig.\,\ref{Fig1}).

The observed redshift of the ICM is $z_{\rm ICM} = 0.2973{\pm}0.0002$, consistent with that of the host galaxy, indicating only a modest line-of-sight bulk velocity. 
The systemic redshift of the host galaxy, measured by the Hubble Space Telescope using ultraviolet absorption lines (such as Ly$\alpha$, Ly$\beta$ and C\,\textsc{iv}), is $z_{\rm abs} = 0.29708{\pm}0.00003$ (ref. \cite{Bahcall1992}), consistent with that measured by optical emission lines of narrow components of the Balmer and forbidden lines (such as [O\,\textsc{iii}] lines from the quasar narrow-line region), $z_{\rm emi} = 0.2972{\pm}0.0002$ (ref. \cite{Shapovalova2016}). 
Thus, the line-of-sight bulk velocity was measured to be $v_{\rm bulk} = 51{\pm}46$~km~s$^{-1}$ relative to the host galaxy. 
This is comparable with or smaller than the bulk velocities measured for other nearby non-merging clusters ($v_{\rm bulk} \lesssim$ 300~km~s$^{-1}$; Cygnus A \cite{Majumder2026}, Perseus \cite{Hitomi2016,XRISM26_Perseus}, Abell 2029 \cite{XRISM25_Abell2029_I,XRISM25_Abell2029_II}, Hydra A \cite{Rose2025}, Virgo \cite{XRISM26_Virgo}, Ophiuchus \cite{Fujita2025} and the Centaurus cluster \cite{XRISM25_Centaurus}).

\begin{figure}[t]
    \centering
    \includegraphics[width=\textwidth]{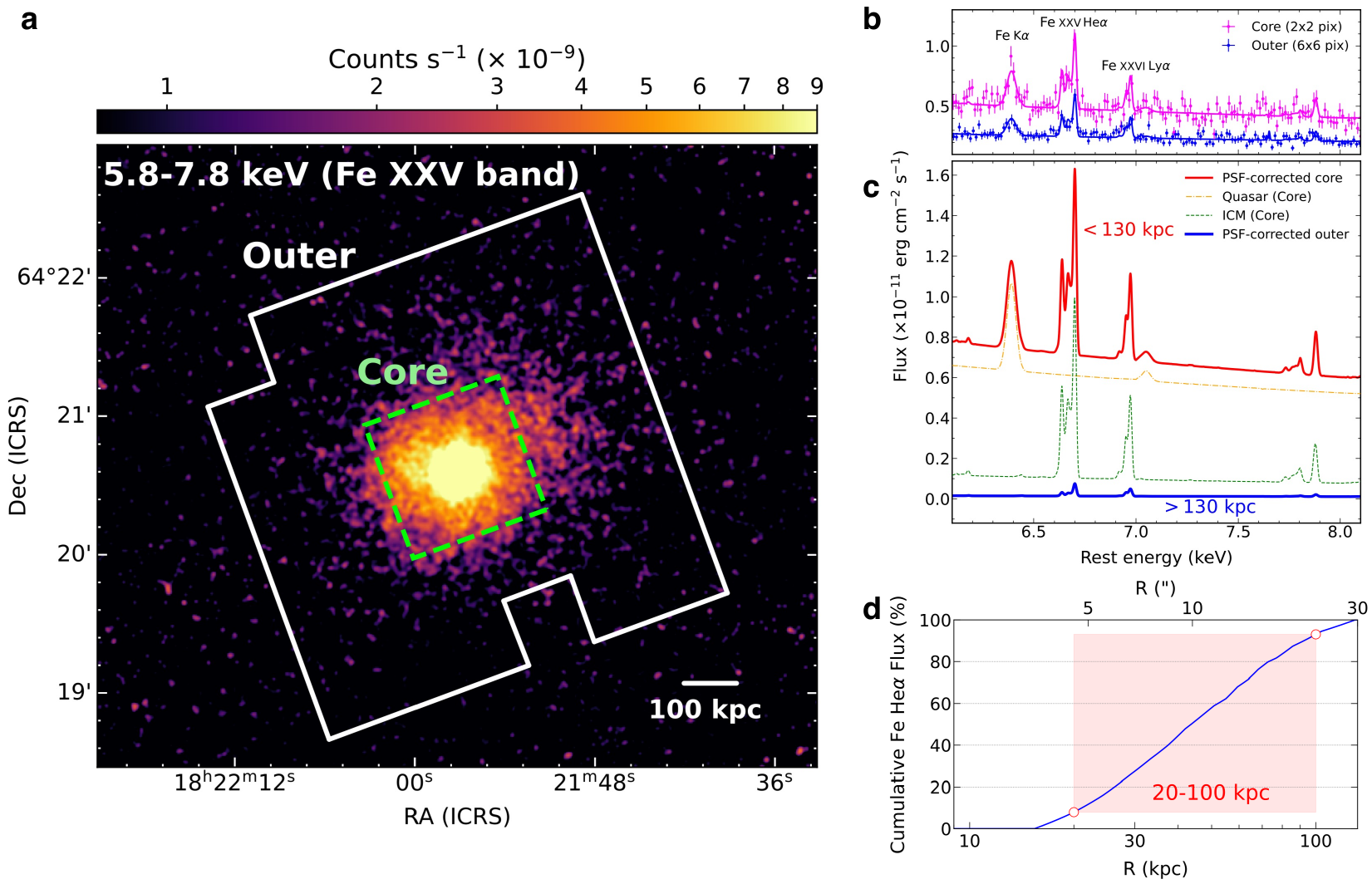}
    \vspace{0.03 cm}
    \caption{{\label{Fig2}\bf Spatial distribution and spectral decomposition of Fe line emission.}
    {\bf a,}
    Background-subtracted Chandra image in the 4.5--6.0~keV band (rest frame 5.8--7.8~keV) highlighting Fe\,\textsc{xxv} He$\alpha$ emission. 
    The green dashed square and white solid line show the central $2 \times 2$~px$^2$ and outer pixels of the Resolve detectors. 
    {\bf b,} Resolve spectra from the core (magenta) and outer (blue) regions with $1\sigma$ uncertainties. Solid curves mark the best-fitting models. 
    {\bf c,} Red and blue curves show the models corrected with the point spread function from the core and outer emission, respectively. 
    The dashed orange and green curves represent the quasar and ICM components in the core. 
    {\bf d,} Cumulative plot of the Fe\,\textsc{xxv} He$\alpha$ line flux as a function of projected radius. 
    The fraction in the 20--100~kpc region is highlighted in red. 
    Dec., declination; ICRS, International Celestial Reference System; PSF, point spread function; RA, right ascension.
    }
\end{figure}

To determine the spatial origin of the highly ionized Fe lines, we extracted XRISM spectra from the core region of the central $2 \times 2$~px$^2$ (or $1' \times 1'$) and the outer region of the other pixels. 
These regions are overlaid on the Chandra rest-frame 5.8--7.8-keV image of H1821+643 (Fig.\,\ref{Fig2}a). 
The spectra from the two regions are almost identical and both contain the neutral Fe~K$\alpha$ fluorescent line from the cold gas around the quasar (Fig.\,\ref{Fig2}b). 
This similarity arises from photon leakage due to the moderate angular resolution (half-power $\sim$ 1.3$'$), as characterized by the point spread function. 
This leakage especially contaminates the outer region due to emission from the bright core, including the quasar. 
To measure the emission from the ICM, we simultaneously fitted these two spectra with the core (quasar plus ICM) and outer (ICM) components by taking into account the appropriate weights, using what is known as a spatial–spectral mixing analysis \cite{Hitomi2018_SSM}. 
As shown in Fig.\,\ref{Fig2}c, the broad emission lines (Fe\,\textsc{xxv} He$\alpha$ and Fe\,\textsc{xxvi} Ly$\alpha$) originate primarily from the ICM in the core region around the quasar, corresponding to projected spatial scales of $\lesssim$130~kpc with a velocity dispersion $\sigma_{v} = 281^{+26}_{-29}$~km~s$^{-1}$ and a temperature $kT = 6.0^{+0.4}_{-0.3}$~keV. 
Less than 10\% of the ICM emission line originated from the outer region, for which we fixed $\sigma_{v}$ to that of the core because of limited statistics and found $kT = 8.2^{+2.8}_{-1.6}$~keV (Methods).

To better constrain the effective spatial scale of the bright core region, we referred to a previous Chandra study \cite{Russell2024} (Methods). 
Using the ICM flux, temperature and metallicity, we estimated the flux of the highly ionized Fe lines. 
Based on the spatial distributions of these parameters from the literature \cite{Russell2024}, we calculated the line fluxes at different radii and constructed a cumulative profile of the Fe\,\textsc{xxv} He$\alpha$ flux in the core region (Fig.\,\ref{Fig2}d).
Over 90\% of the Fe\,\textsc{xxv} He$\alpha$ flux originated from the 20--100~kpc region. 
The line flux from the innermost $\lesssim$4$''$ ($\lesssim$18~kpc) region is negligible, consistent with the nuclear Chandra spectrum without Fe\,\textsc{xxv} and Fe\,\textsc{xxvi} lines (Extended Data Fig.\,\ref{Ext_Fig3}). 
This rules out substantial contamination from photoionized lines near the quasar. 
These results demonstrate that, within the core region, nearly all of the lines originated from the 20--100~kpc region.

If the large velocity dispersion in H1821+643 arises from turbulence or radial flows, the non-thermal-to-thermal energy ratio $f_{\rm nth} = 8.4^{+1.6}_{-1.8}\%$, which is mostly within the 20--100-kpc region (Methods). 
This value is higher than the fractions observed within comparable scales (${\lesssim}100$~kpc) in other clusters observed by XRISM and Hitomi, where $f_{\rm nth} \approx 1$--5\% in clusters with lower AGN luminosities (Supplementary Figs.\,\ref{Sup_Fig1}--\ref{Sup_Fig2}).

\begin{figure}[t]
    \centering
    \includegraphics[width=\textwidth]{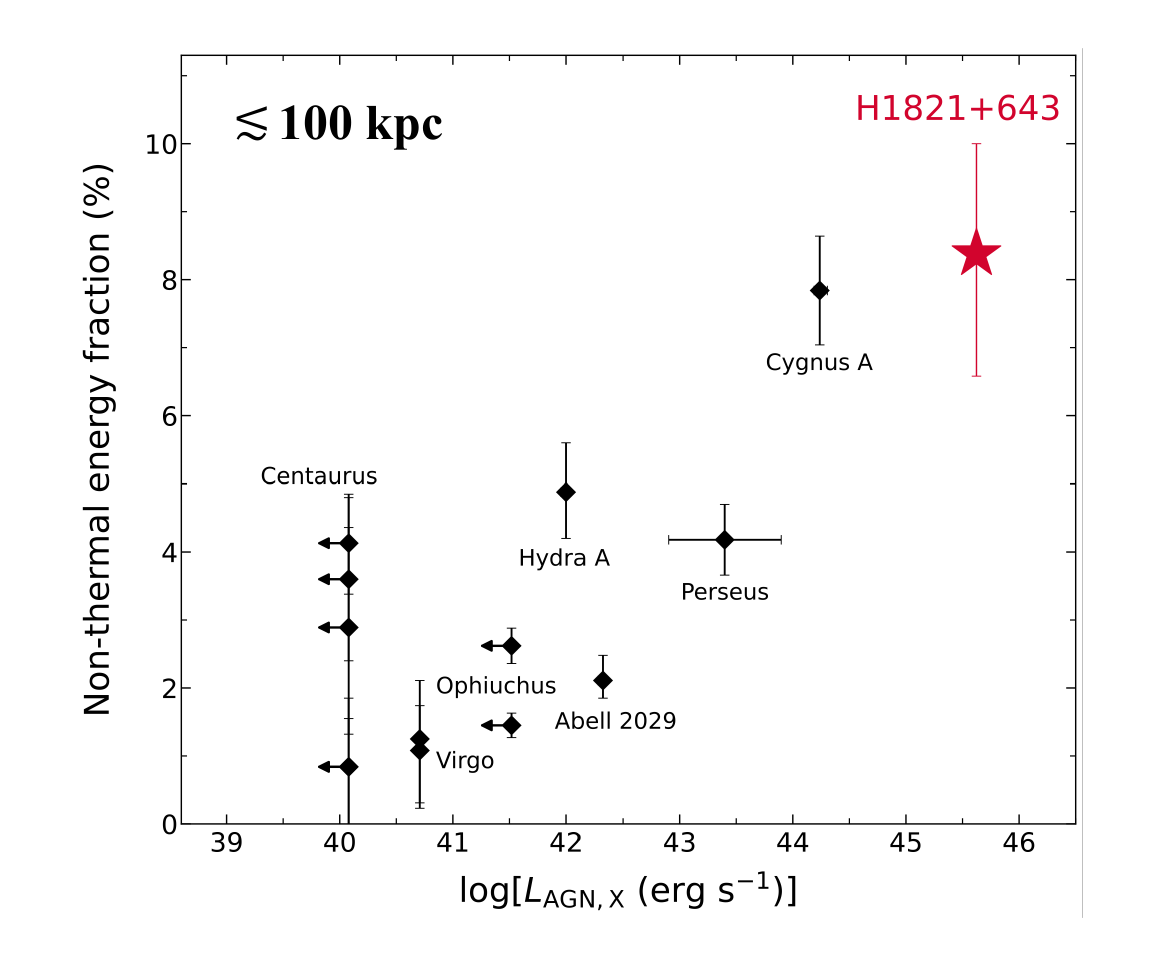}
    \vspace{0.3 cm}
    \caption{
    {\label{Fig3}\bf Non-thermal energy fraction within $\lesssim$100~kpc versus AGN luminosity.}
    The non-thermal-to-thermal energy fraction ($f_{\rm nth}$), measured within $\lesssim$100~kpc of the hot ($>$2 keV) ICM, is plotted against the 2--10-keV AGN luminosity for clusters observed with XRISM and Hitomi.
    Red star marks H1821+643, while black diamonds denote other clusters.
    Error bars indicate $1\sigma$ uncertainties, and arrows denote $3\sigma$ upper limits.
    H1821+643, despite hosting a radio-quiet quasar, exhibits the most vigorous gas motions (turbulence and velocity shear) and the brightest X-ray AGN.
    }
\end{figure}

Figure~\ref{Fig3} compares $f_{\rm nth}$ with the intrinsic, absorption-corrected 2--10-keV X-ray luminosity of AGNs ($L_{\rm AGN,X}$). 
The figure indicates that $f_{\rm nth}$ increases with AGN luminosity. 
Notably, Cygnus A is radio-loud \cite{Majumder2026}, whereas H1821+643 ($L_{\rm AGN,X} \sim 4.2\times10^{45}$~erg~s$^{-1}$) is radio-quiet yet shows the highest $f_{\rm nth}$, which indicates that powerful quasars can drive large velocity dispersions regardless of the presence of jets.
Although it is difficult to distinguish whether the velocity dispersion arises from turbulence or velocity shear---including motions such as sloshing or radial winds---the ICM motions in this quasar–host system are exceptionally vigorous.


H1821+643 hosts a giant radio halo extending over $\sim$1.1~Mpc (ref. \cite{Bonafede2014}). 
However, particle acceleration through violent mergers is probably not the origin of the halo because the cool core has survived. 
We note that relaxed cool-core clusters, such as Perseus and Abell~2029, also host large radio haloes \cite{van_Weeren2024,Govoni2009}. 
This indicates that even minor mergers can create haloes \cite{Bonafede2014}. 
The extended halo in H1821+643 may have originated from minor mergers unrelated to the central ICM motions.

In the core region, Chandra images of H1821+643 reveal a cold front and some asymmetries, potentially indicating sloshing in the ICM \cite{Russell2010,Russell2024}. 
Current observations do not allow a definitive assessment of its contribution to the measured linewidth. 
If sloshing were occurring, it could contribute to broadening, but the total velocity range implied by the observed $\sim$300~km~s$^{-1}$ dispersion would require several components spanning $\Delta v_{\rm bulk} \sim 600$~km~s$^{-1}$ (Methods). 
For comparison, in the Centaurus cluster, the line-of-sight bulk velocities range from $-$130~km~s$^{-1}$ to $-$310~km~s$^{-1}$, corresponding to $\Delta v_{\rm bulk} \sim 180$~km~s$^{-1}$ (ref. \cite{XRISM25_Centaurus}). 
Given the near-zero net bulk velocity and the extraordinarily large velocity range in H1821+643, such a highly symmetric configuration appears unlikely.

\begin{figure}[t]
    \centering
    \includegraphics[width=0.95\textwidth]{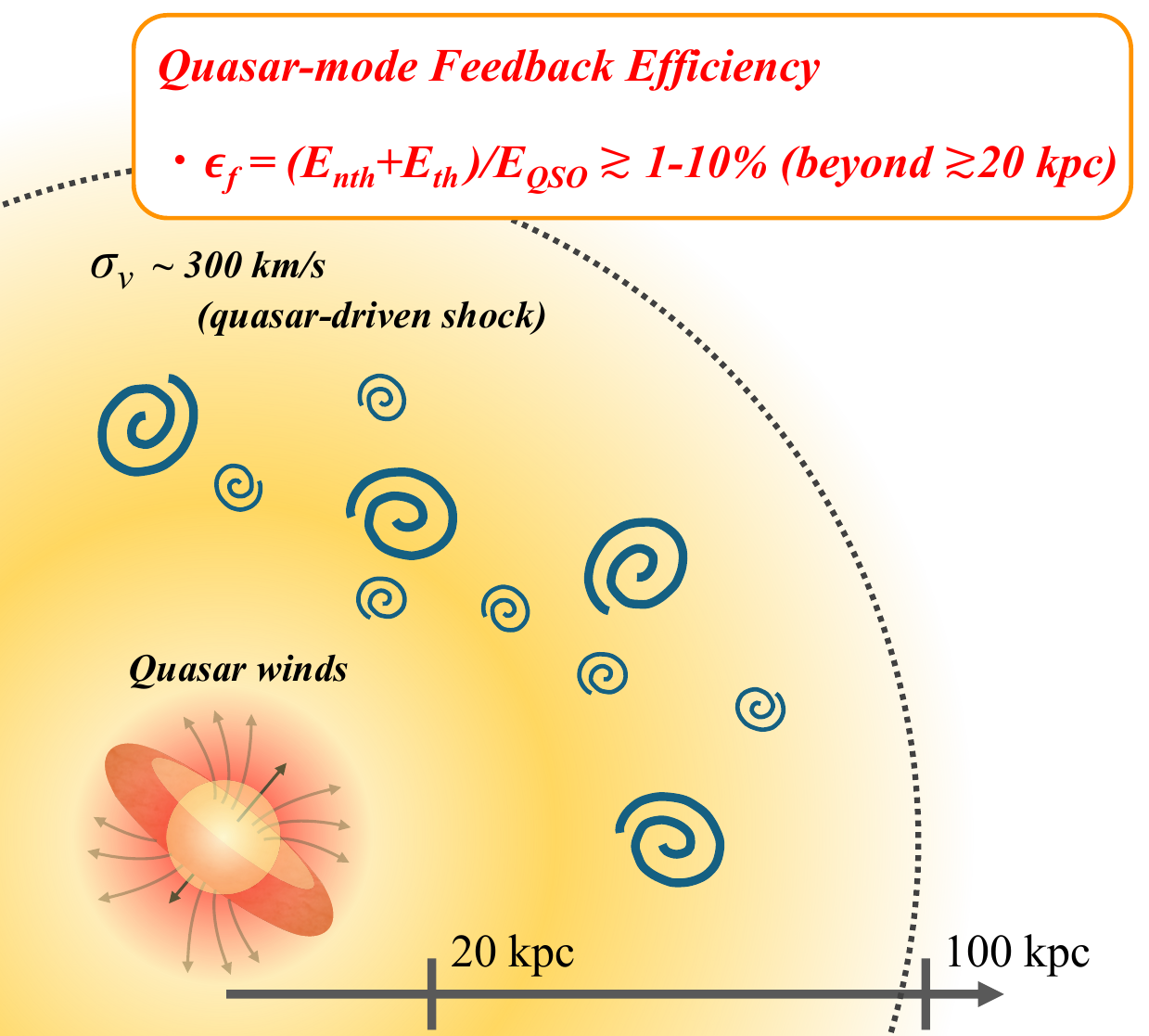}
    \vspace{0.5 cm}
    \caption{
    {\label{Fig4}\bf Schematic of energy injection into the ICM beyond galactic scales via quasar-mode feedback.}
    On galaxy scales ($<$20~kpc), the luminous quasar (yellow sphere) inflates a hot gas region (red sphere), thereby launching energetic winds (grey arrows) along the polar axis (black arrows) perpendicular to the accreting gas ring. 
    At cluster core scales of 20--100~kpc, the winds expand and drive a forward shock wave that propagates supersonically through the surrounding ICM (dashed circle). The quasar-driven shock amplifies the turbulence and velocity shear (spiral curves), thereby producing velocity dispersions of $\sim$280~km~s$^{-1}$ (or $\sim$300~km~s$^{-1}$ if resonance scattering is negligible). 
    The resulting large velocity dispersion and thermal structure of the ICM (large yellow sphere) carry the imprint of quasar energy injection and provide strong constraints on the efficiency of quasar-mode feedback.
    }
\end{figure}

A plausible driver of non-thermal energy injection is powerful quasar-driven winds. 
As these winds expand, they can drive a forward shock that propagates supersonically through the surrounding ICM with a temperature of $\sim$6~keV (its speed of sound is 1,260~km~s$^{-1}$), hereafter referred to as the ‘quasar-driven shock’ (Fig.\,\ref{Fig4}). 
Because of the high temperature, the shock will probably weaken rapidly. 
If the shock front has propagated beyond $\gtrsim$100~kpc, its low surface brightness would make it difficult to detect, as predicted by mock Chandra simulations of weak shocks with typical Mach numbers of the order of $\sim$1.1 (ref. \cite{Prunier2025c}). 
This is consistent with the absence of obvious shock features in deep Chandra imaging of H1821+643 \cite{Russell2024}. 
Instead, the vigorous gas motions expected to be generated behind the shock may leave observable imprints, as indicated by our XRISM measurements. 
The measured linewidth does not trace the shock speed itself, but the velocity dispersion of the post-shock gas, which is expected to be substantially smaller for such weak shocks. 
Within 20--100~kpc, the ICM mass is estimated as $M_{\rm ICM} \simeq 1.6 \times 10^{12} M_{\odot}$, based on the volume-weighted electron density ($\bar{n}_{e} = 0.013$~cm$^{-3}$) measured with Chandra \cite{Russell2024} (Methods). 
The corresponding non-thermal energy, for $\sigma_{v} = 281^{+26}_{-29}$~km~s$^{-1}$, is $E_{\rm nth}$ = $(3/2)M_{\rm ICM}\sigma_{v}^2 \simeq 4 \times 10^{60}$~erg, an order of magnitude above the mechanical energy from X-ray cavities attributed to jets ($3 \times 10^{59}$~erg) \cite{Russell2024}.

Owing to the high temperature of the ICM, the shock is expected to be weak. 
In this scenario, the required shock expansion energy injected by the quasar is approximately the product of the surrounding ICM pressure and the volume inside the shock \cite{McNamara2005}. 
This energy is comparable with the thermal energy of the ICM within the shock. 
Assuming that the shock radius is $\sim$100~kpc, which is comparable with the cooling radius (Supplementary Table\,\ref{Sup_Tab1}), the enclosed thermal energy $E_{\rm th} \sim 4 \times 10^{61}$~erg. 
For comparison, we define the total energy potentially injected by the quasar as $E_{\rm QSO} \lesssim \eta M_{\rm BH} c^2 \sim 4 \times 10^{62}$~erg, corresponding to the maximum energy available from the central supermassive black hole with radiative efficiency $\eta \sim 0.1$ and $c$ being the speed of light, which indicates that the black hole could, in principle, supply the energy required to drive the observed turbulence and velocity shear. 
Comparing these values implies that the quasar-mode feedback in H1821+643 operates with an efficiency $\epsilon_{\rm f} = (E_{\rm nth} + E_{\rm th})/E_{\rm QSO} \gtrsim 10$\% beyond galactic scales ($\gtrsim$20~kpc). 
However, this estimate strongly depends on the assumed shock radius. 
Alternatively, considering only the non-thermal energy supplied by the current quasar activity provides a conservative lower limit on the feedback efficiency, $\epsilon_{\rm f} \gtrsim E_{\rm nth}/E_{\rm QSO} \gtrsim 1$\%. 
This level of coupling, $\epsilon_{\rm f} \gtrsim 1$--10\%, is far higher than previous estimates of $\lesssim$0.01\% from similar-scale ionized winds in other quasars \cite{Lacy2019,Venturi2023} but consistent with the energy requirements of recent cosmological hydrodynamical simulations \cite{Husko2024,Husko2026,Schaye2026}. 
Taken together, these observations provide compelling evidence that quasar-mode feedback can substantially influence the surrounding ICM over 20--100~kpc scales.

From a theoretical perspective, state-of-the-art cosmological simulations adopt feedback efficiencies of comparable magnitudes. 
For example, the COLIBRE simulation \cite{Schaye2026} uses thermal coupling efficiencies $\epsilon_{\rm f}\simeq 5$--10\% in its fiducial model and somewhat lower values ($\epsilon_{\rm th}\simeq 2$--3\%) supplemented by comparable kinetic jet power in its hybrid implementation \cite{Husko2026}.
The IllustrisTNG simulations similarly show that quasar-mode (thermal) feedback dominates at high accretion rates and redshift, whereas jet-mode (kinetic) feedback governs the heating and stirring of the ICM in cluster cores at low redshift. 
Our inferred feedback efficiency of $\gtrsim$1--10\% lies within the ranges assumed in these models. 
However, we note that feedback efficiencies in cosmological simulations are defined at the subgrid injection level and depend on the assumed coupling scale and timescale and are, therefore, not strictly identical to our cumulative, radius-integrated estimate.
Thus, the comparison is qualitative in terms of overall energetics and indicates broad consistency with feedback prescriptions used in large-scale simulations, at the level of order $\epsilon_{\rm f} \gtrsim 1$--10\%.

Notably, these results were obtained in the regime of sub-Eddington accretion, with an Eddington ratio $\lambda_{\rm Edd}\sim 0.3$--0.6, and subsonic turbulence (Mach number $\sim 0.4$), consistent with the ‘gentle’ feedback mode expected in cool-core clusters.
Radio observations reveal that H1821+643 hosts a low-power FR I-like jet \cite{Blundell2001}, with mechanical energy from X-ray cavities estimated at $3 \times 10^{59}$~erg, well below the jet powers predicted in the COLIBRE hybrid model \cite{Husko2026} and on the lower side of injected powers seen in TNG-Cluster cavity populations \cite{Prunier2025a}. 
This discrepancy underscores a key tension: simulations tend to predict jet-mode dominance in massive haloes at low redshift, yet H1821+643 seems to be governed by quasar-mode feedback, with radiative winds and bubble expansion probably driving the observed turbulence and velocity shear. 
The presence of such a luminous, radiatively efficient supermassive black hole at the centre of a massive cluster at $z \simeq 0.3$ is, therefore, highly unusual and offers a rare opportunity in the local Universe to directly study how quasar-mode feedback couples to the ICM. 
H1821+643 can, thus, be viewed as a nearby analogue of the powerful high-redshift quasars thought to regulate galaxy growth, preheat protoclusters and enrich the intergalactic medium.

\clearpage

\clearpage
\backmatter



\section*{Methods}
\bmhead{XRISM observations and data extraction}
This study presents the first observation performed under the XRISM guest observer programme (cycle 1). XRISM observed H1821+643 between 4 and 10 September 2024, with a total net exposure of 283.7~ks (observation ID 201030010). 
The spacecraft was equipped with the Resolve instrument, a high-energy-resolution X-ray microcalorimeter spectrometer coupled with an X-ray mirror assembly providing a $3' \times 3'$ field of view \cite{Tashiro2025,Ishisaki2025,Kelley2025}. 
Resolve achieved an energy resolution of ${<}4.9$~eV full-width at half-maximum at 6~keV (ref. \cite{Ishisaki2025}), thereby allowing precise measurements of line centroids and widths.

The Resolve observation was performed with an open filter and closed gate valve, which restricted the effective band-pass to $E \approx 1.7$--12 keV. 
Data reduction used the pre-pipeline package (v.\texttt{005\_002.20Jun2024\_Build8.012}), the standard pipeline script (03.00.013.009) and CALDB 20240815. 
Good-time-interval filtering excluded periods affected by the eclipse of the Earth, sunlit Earth limb (${<}20^\circ$), passages through the South Atlantic anomaly and the first 4,300~s after the 50-mK cooler recycling. 
Events were screened using pixel-to-pixel coincidence to remove multi-pixel cosmic-ray events, with a 300-eV threshold and with the standard energy-dependent rise-time cut to prevent crosstalk \cite{Kilbourne2018,Mochizuki2025}. 
Only high-resolution primary events, which provide optimal spectral performance, were used for the scientific analysis. 
Low-resolution secondary events, which potentially contain spurious or degraded signals, were excluded to preserve data quality. 
This is especially important for this faint source. 
Pixel 12 (calibration pixel) and pixel 27 (prone to unpredictable gain jumps) were omitted from the analysis \cite{Eckart2025,Porter2025}. 
Given the low count rate of the target ($\lesssim$0.03 counts per second per pixel), residual contamination from crosstalk or high-count-rate effects was negligible \cite{Ishisaki2018,Mizumoto2025}.

Time-dependent gain calibration was performed using onboard $^{55}$Fe sources on the filter wheel. 
Fourteen fiducial measurements were taken at the start, end and at several intermediate intervals during the observation. 
The Mn K$\alpha$ line complex was fitted per pixel with multi-component models to correct for gain drifts, which yielded an energy offset below 0.1~eV and a full-width at half-maximum of $4.37{\pm}0.02$~eV. 
The statistical uncertainty was far smaller than the overall systematic uncertainty of the energy scale ($<$0.3~eV, corresponding to $<$15~km~s$^{-1}$ in velocity \cite{Eckart2025}), indicating precise calibration. 
These gain drift corrections and systematic uncertainties were fully propagated into the emission-line centroid and width measurements to provide robust constraints on bulk motions and turbulence.
The heliocentric correction was negligible (approximately $-1$~km~s$^{-1}$).

Redistribution matrix files (RMFs) were generated with \texttt{rslmkrmf} using the extra-large option (X) and the parameter file \texttt{xa\_rsl\_rmfparam\_20190101v006.fits.gz} to incorporate the Gaussian core, exponential low-energy tail, escape peaks, silicon fluorescence and electron-loss continuum. 
Auxiliary response files (ARFs) were generated with \texttt{xaarfgen} using the Chandra 0.5--7.0~keV image of the cluster from ref. \cite{Russell2024} to account for its spatially extended emission. 
Non-X-ray background (NXB) spectra were generated from the stacked night-Earth database using \texttt{rslnxbgen} by applying the same good time intervals as for the source. 
The baseline NXB model comprised a power-law continuum and Gaussian lines representing instrumental features from Al, Au, Cr, Mn, Fe, Ni and Cu.
The NXB spectra were fitted independently using a diagonal RMF without an ARF. 
The best-fitting parameters were subsequently fixed when modelling the source spectra with the $C$-statistic. 
The contribution of the NXB in the 4--7-keV energy range was negligible (Extended Data Fig.\,\ref{Ext_Fig1}).
\\

\bmhead{Reduction of Hitomi and Chandra data}
To compare the highly ionized Fe\,\textsc{xxv} He$\alpha$ emission, we reproduced the spectrum of the ICM in the Perseus cluster core observed by Hitomi in 2016 \cite{Hitomi2016}. 
Data reduction followed the procedure outlined in ref. \cite{Hitomi2018_SSM} to ensure agreement with the published results. 
RMFs were generated using the \texttt{sxsmkrmf} tool, which incorporated the electron-loss continuum channel into the redistribution to produce an extra-large RMF. 
ARFs were generated using the \texttt{aharfgen} tool. 
Only high-resolution primary events were extracted, and time-dependent gain corrections were applied following equation (A1) of ref. \cite{Hitomi2018_Atomic}. 
To verify reproducibility, the spectrum was binned to at least 1 count per bin and fitted with the same ICM model and a negative Gaussian for the w-line, thereby yielding a velocity dispersion of $\approx$164~km~s$^{-1}$ and a temperature of $kT \simeq 4.0$~keV, consistent with published results \cite{Hitomi2016} (Fig.\,\ref{Fig1}).

For H1821+643, we used archival Chandra ACIS-S data (observation IDs in ref. \cite{Russell2024}), with a total exposure of 582~ks observed between 2019 and 2020. 
The datasets were reprocessed using \texttt{CIAO} v4.17 with \texttt{CALDB} v4.12 \cite{Fruscione2006}. 
A background-subtracted Chandra image in the 4.5--6.0~keV band (corresponding to 5.8--7.8~keV in the rest frame) was extracted to highlight Fe\,\textsc{xxv} He$\alpha$ emission and used to produce Fig.\,\ref{Fig2}.
X-ray spectra were generated using the \texttt{specextract} task, and the resulting spectra, RMFs and ARFs were combined for spectral fitting.
\\

\bmhead{Spectral analysis}
Spectra were binned to at least 1 count per bin for spectral fitting using the $C$-statistic \cite{Cash1979} to preserve information in this faint target, whereas binning appropriate for display was applied for the figures. 
Extended Data Fig.\,\ref{Ext_Fig1} shows the broadband (3--10~keV) spectrum of H1821+643 obtained with Resolve. 
Prominent highly ionized lines, including Fe\,\textsc{xxv} He$\alpha$ and Fe\,\textsc{xxvi} Ly$\alpha$, were detected alongside a neutral Fe K$\alpha$ feature associated with the quasar. 
The ICM emission was described by a collisionally ionized plasma model (\texttt{bapec} model with AtomDB v.3.0.9 in XSPEC v.12.14.1 \cite{Arnaud1996}), with temperature, metallicity, redshift, velocity dispersion and normalization left free unless otherwise noted. 
All abundances are given relative to the \texttt{lpgs} proto-solar values \cite{Lodders2009}.
The unobscured AGN emission was modelled with a \texttt{zpowerlw} continuum plus three Gaussian components for the neutral Fe K lines: Fe K$\alpha_1$ (6.404~keV), Fe K$\alpha_2$ (6.391~keV) and Fe K$\beta$ (7.06~keV), which had relative line intensities of 1, 0.5 and 0.15, respectively. 
For the AGN component, the photon index of ${\approx}1.9$ and 2--10-keV luminosity of ${\approx}4.2 \times 10^{45}$~erg~s$^{-1}$ were consistent with previous measurements with Chandra \cite{Russell2024}. 
The intrinsic width of the neutral Fe lines ($\sigma_{\rm Fe} \approx 21$~eV) probably originated from the broad-line region of the quasar. 
The AGN structure will be discussed in detail in a forthcoming work (Yamada et al., in preparation).

Spectral fitting was performed in the 4.0--7.0~keV band, chosen to capture prominent emission lines, yielding an ICM velocity dispersion $\sigma_v \approx 295$~km~s$^{-1}$, $z_{\rm ICM} \approx 0.297$, $kT \sim 6$~keV with metallicity ${\sim}0.6 Z_{\odot}$ ($Z_{\odot}$ being the solar metallicity). 
As He-like and H-like lines can trace plasma at different temperatures or regions, we first examined the broadened He-like Fe\,\textsc{xxv} He$\alpha$ line identified in Fig.\,\ref{Fig1}. 
Excluding the H-like Fe\,\textsc{xxvi} band and fixing the metallicity at $0.6 Z_{\odot}$ (from Chandra results) yielded $\sigma_v = 285^{+22}_{-42}$~km~s$^{-1}$, $z_{\rm ICM} = 0.2975{\pm}0.0002$ and $kT = 6.2{\pm}0.5$~keV. 
When the H-like line was included and the full 4--7~keV band fitted, we obtained a consistent velocity dispersion $\sigma_v = 295^{+34}_{-28}$~km~s$^{-1}$. 
A two-temperature model was tested, but the $C$-statistic showed no significant improvement, consistent with analyses of Perseus and other clusters; thus, we adopted a single-temperature model.

To assess the impact of resonance scattering, the Fe\,\textsc{xxv} He$\alpha$ complex around 5.10--5.18~keV (rest frame 6.62--6.72~keV) was masked and the resulting fit yielded consistent results ($\sigma_v \approx 302$~km~s$^{-1}$), although the resonance (w) line appeared weaker than predicted (Extended Data Fig.\,\ref{Ext_Fig2}). 
This is probably attributable to the slightly lower temperature inferred when the Fe\,\textsc{xxv} He$\alpha$ complex was excluded ($kT \sim 5.8^{+0.5}_{-0.4}$~keV), which is still consistent within the uncertainties of the other fits and to the potential effect of resonance scattering. 
A negative Gaussian absorption component ({\tt zgabs}) was included at the w-line to account for this effect, resulting in $\sigma_v = 283^{+26}_{-29}$~km~s$^{-1}$, $kT = 6.2^{+0.3}_{-0.2}$~keV, $z_{\rm ICM} = 0.2973{\pm}0.0002$ and metallicity $Z = 0.60{\pm}0.04 Z_{\odot}$, which we adopted as the fiducial model. 
The linewidth ${\gtrsim}8.4$~eV (allowing for variations within 10~eV) and the line depth was $0.96^{+0.25}_{-0.24}$~keV, indicating that the w-line flux was suppressed by $\sim$30\%. 
This corresponds to a modest optical depth $\tau \approx 1$, which is close to the value observed for the Perseus cluster ($<$30~kpc) \cite{Hitomi2018_RS}.
Therefore, although resonance scattering may be effective and incompatible with high turbulence, it is thought to occur in the central region of H1821+643. 
This is because resonance scattering is usually observed only in central regions ($\lesssim$30~kpc) like the Perseus cluster, and it is unlikely that $\tau \approx 1$ would occur in the 20--100-kpc region. 
The improvement in the $C$-statistic was small ($\Delta C \simeq 7$), implying that the effect is not tightly constrained. 
Using the Akaike information criterion \cite{Akaike1974} to assess the statistical significance, we found that the presence of resonance scattering was not unambiguously required, with a null hypothesis probability of $P_{\rm AIC} \approx 22$\%. 
Therefore, the derived velocity dispersion including the negative Gaussian component ($\sigma_v = 283^{+26}_{-29}$~km~s$^{-1}$) is regarded as a conservative lower limit, whereas the fit without this component ($\sigma_v = 295^{+34}_{-28}$~km~s$^{-1}$) provides an approximate upper bound; the data do not permit a more precise determination. 
Although the fit was restricted to the 4--7-keV band, the model also reproduced the full 3--10-keV spectrum of H1821+643 (Extended Data Fig.\,\ref{Ext_Fig1}).
\\

\bmhead{Testing two-component models}
We tested alternative spectral models to assess whether several plasma temperatures or velocity components could contribute to the observed line broadening. 
A two-temperature model provided no significant improvement ($\Delta C < 2$ for five more \texttt{bapec} parameters), probably because soft X-ray lines are difficult to detect in this $z \sim 0.3$ target against the AGN continuum.
Therefore, we adopted a single-temperature (1T) model as a conservative estimate and ensured that comparisons with other clusters were consistently made using 1T model values.

We also explored a simplified two-velocity-component model, which shifted the redshift of two identical plasma components to blue and red. 
The observed linewidth of H1821+643 ($\sigma_v \sim 300$~km~s$^{-1}$) corresponds to a separate bulk motion of $\sim$600~km~s$^{-1}$. 
Assuming velocity dispersions typical of other clusters such as Perseus ($\sigma_v \lesssim 160$~km~s$^{-1}$) yielded $\Delta v_{\rm bulk} \gtrsim 540$~km~s$^{-1}$, although the fitting improvement was minimal ($\Delta C \sim 1$). 
Although highly idealized, this indicates that a bulk-dominated interpretation would require an unusually large velocity separation ($\sim$600~km~s$^{-1}$) compared with typical clusters, even when accounting for the velocity dispersion expected from normal ICM motions.
\\

\bmhead{Spatial-spectral mixing analysis}
To separate the core ($2 \times 2$ px$^2$) from the outer 30~px, we performed a spatial–spectral mixing analysis \cite{Hitomi2018_SSM} (Fig.\,\ref{Fig2}).
We assumed that the X-ray emission followed the Chandra 0.5--7.0-keV image of H1821+643 described in the data reduction section. 
Ray-tracing simulations were conducted to generate ARFs that account for photon leakage between regions. 
The core emission was modelled by the ICM (\texttt{bapec}) and the quasar emission (\texttt{zpowerlw} $+$ $3$ $\times$ \texttt{zgauss}), whereas the outer emission was modelled by the ICM alone. 
Given the temperature distribution measured with Chandra, the core and outer ICM temperatures were left free. 
NXB spectra were generated separately for each region. 
Although the normalization of the NXB components was a free parameter, their ratio matched the number of detector pixels, so the flux ratio of the NXB components was fixed according to the pixel areas. 
The spectra extracted from the two regions were fitted simultaneously using their respective ARF and RMF files, while we kept the NXB models fixed. 
The metallicity of the outer ICM was fixed at $0.3 Z_{\odot}$ from Chandra measurements \cite{Russell2024}, as the faint emission does not allow meaningful constraints. 
Varying the outer metallicity had a negligible impact on the bright core parameters.

To verify the robustness of the fit, the outer component parameters (except the metallicity) were first allowed to vary freely. 
This resulted in best-fitting values $z \sim 0.2988$, $\sigma_v \sim 0$~km~s$^{-1}$ and $kT \sim 6.4$~keV for the outer, requiring an unrealistically large bulk velocity with small dispersion. 
Next, we fitted both regions with a common redshift, yielding $z = 0.2973^{+0.0003}_{-0.0001}$, $\sigma_v = 287^{+35}_{-33}$~km~s$^{-1}$ and $kT = 6.0^{+0.4}_{-0.3}$~keV for the core, whereas the outer parameters remained poorly constrained, with $\sigma_v \gtrsim 66$~km~s$^{-1}$ and $kT = 7.2^{+4.8}_{-1.0}$~keV, inconsistent with independent Chandra measurements ($kT > 8$~keV; ref. \cite{Russell2024}).

Finally, we adopted a model with the outer region parameters $z$ and $\sigma_v$ linked to the core values, thereby prioritizing its small contribution and consistency with spatially resolved observations. 
In this final model, the broad Fe\,\textsc{xxv} He$\alpha$ and Fe\,\textsc{xxvi} Ly$\alpha$ lines arose predominantly from the core ICM ($\lesssim$130~kpc), with $z = 0.2973{\pm}0.0002$, $\sigma_v = 281^{+26}_{-29}$~km~s$^{-1}$, $kT = 6.0^{+0.4}_{-0.3}$~keV and $Z = 0.73^{+0.06}_{-0.03} Z_{\odot}$. 
The outer region contributed $<$10\% of the ICM line emission, with $kT = 8.2^{+2.8}_{-1.6}$~keV, and freeing the outer $z$ did not significantly affect the core parameters (Fig.\,\ref{Fig2}c).
\\

\bmhead{Cumulative Fe\,\textsc{xxv} plot and nuclear Chandra spectrum}
To better constrain the effective spatial scale of the bright core, we used published Chandra measurements of the ICM flux, temperature and metallicity in annuli around H1821+643 (ref. \cite{Russell2024}). 
Using these parameters (\texttt{apec}), we calculated the expected Fe\,\textsc{xxv} line flux in each region and constructed a cumulative profile of the Fe\,\textsc{xxv} He$\alpha$ flux (Fig.\,\ref{Fig2}d). 
The 20--100-kpc region, where the ICM temperature spanned ${\sim}3$ to 9~keV, contributed over 90\% of the Fe\,\textsc{xxv} flux. 
Chandra measurements gave a core metallicity of ${\sim}0.6 Z_{\odot}$ and an outer metallicity of ${\sim}0.3 Z_{\odot}$, consistent with values from the spatial–spectral mixing analysis. 
The volume-weighted electron density in the 20--100-kpc shell $\bar{n}_{e} \approx 0.013~\mathrm{cm^{-3}}$, from which the total gas mass can be estimated using $M_{\rm gas} = \mu_e m_p \bar{n}_{e} V$, where $\mu_e \approx 1.17$ is the mean molecular weight per free electron, $m_p$ is the proton mass and $V$ is the volume. 
This yielded a total gas mass $M_{\rm gas} \sim 1.6 \times 10^{12}~M_\odot$ in this region.

The cumulative profile shows that the Fe\,\textsc{xxv} contribution from the $<$20-kpc region was negligible.
Consistently, the nuclear Chandra spectrum extracted from the innermost ${\lesssim}4''$ showed no detectable highly ionized lines, as only the neutral Fe K$\alpha$ line was detected (Extended Data Fig.\,\ref{Ext_Fig3}). 
Adding a Gaussian component for Fe K$\alpha$ (6.4~keV) to a power-law continuum improved the $C$-statistic ($\Delta C = 56$), whereas including the Fe\,\textsc{xxv} line (6.7~keV) and Fe\,\textsc{xxvi} line (7.0~keV) produced $\Delta C < 1$, indicating no statistically significant detection. 
This confirmed that there was negligible contamination from photoionized plasma near the quasar. 
The absence of the Fe\,\textsc{xxv} and Fe\,\textsc{xxvi} lines from the innermost region was also expected for a plasma with $kT \lesssim 2$~keV. 
These results are consistent with previous Chandra analyses \cite{Russell2024}.
\\

\bmhead{Acknowledgements}
We are grateful to Stefano Ettori, Fabio Gastaldello and the anonymous reviewer(s) for their constructive comments, which have improved this manuscript.
We thank S. Kimura for insightful discussions and helpful comments that improved the interpretation of the quasar-driven ICM turbulence and velocity shear in this work.
\\

\bmhead{Author contributions}
Satoshi Yamada and S.U. conceived the XRISM observation campaign and analysed the XRISM, Hitomi and Chandra data. 
Satoshi Yamada and H.N. led the writing of the AGN and quasar-related sections. 
Satoshi Yamada, S.U. and Y.F. wrote the cluster-related sections. 
M.M. reduced and evaluated the Resolve data. 
K.N. supported the discussions and interpretations based on state-of-the-art cosmological simulations. 
C.R., S.O, T.K., Shinya Yamada, Y.T. and Y.U. contributed to the assessment of the data quality and X-ray spectral fitting and provided critical comments on the overall paper. 
The final version of the article was prepared and approved by the corresponding authors Satoshi Yamada, S.U., H.N. and Y.F., who coordinated all aspects of the project.
\\

\bmhead{Funding}
This work was supported by the JSPS (KAKENHI Grant Nos. 
22K20391, 23K13154 and 26K17187 to Satoshi Yamada, 
25K23398 and 26K00741 to S.U., 
19K21884, 20H01947, 20H01941, 23K20239, 24K00672 and 25H00660 to H.N., 
22H00158, 23H04899 and 25H00672 to Y.F., 
21K13958 to M.M., 
22K21349, 24H00002, 24H00241 and 25K01032 to K.N., 
24K17104 to S.O., 
23K13153 and 24K00673 to T.K.,
22H01272, 23K22543 and 24K00672 to Shinya Yamada, and 
20H01946, 21H04496 and 25H00671 to Y.U.). 
Satoshi Yamada and T.K. acknowledge support from the RIKEN SPDR Program. 
S.U. acknowledges support from the Program for Forming Japan’s Peak Research Universities (Grant No. JPJS00420230006). 
K.N. acknowledges support from the Kavli IPMU, the World Premier Research Centre Initiative, UTIAS and the University of Tokyo. 
C.R. acknowledges support from an SNSF Consolidator grant (No. F01–13252), a Fondecyt Regular grant (No. 1230345), an ANID BASAL project (No. FB210003) and the China–Chile joint research fund. 
Y.U. acknowledges support from the Kyoto University Foundation.
\\

\bmhead{Data availability}
The observational data used in this study are publicly available via the High Energy Astrophysics Science Archive Research Center (HEASARC) at NASA at \url{https://heasarc.gsfc.nasa.gov/}. 
XRISM data are available under observation ID 201030010. 
Hitomi data are available under observation IDs 10040010, 10040020, 10040030, 10040040, 10040050 and 10040060. 
Chandra data are available under observation IDs 21558--21561, 22103--22109, 23053--23054, 23211, 23239--23240, 23319, 23339, 24612, 24639, 24641 and 24661. 
The atomic databases used in the spectral analysis, such as AtomDB, are available online at \url{http://www.atomdb.org/}.
Source data are provided with this paper.

\bmhead{Code availability}
The data reduction for this study was performed using publicly available software provided by HEASARC
(\url{https://heasarc.gsfc.nasa.gov/docs/software/heasoft}). 
Spectral fitting was carried out using the XSPEC package, which is freely accessible online (\url{https://heasarc.gsfc.nasa.gov/xanadu/xspec}).
\\

\bmhead{Competing interests}
The authors declare no competing interests.
\\

\backmatter
\setcounter{figure}{0}

\begin{figure*}
    \captionsetup{name=Extended Data Fig.}
    \centering
    \includegraphics[width=0.9\textwidth]{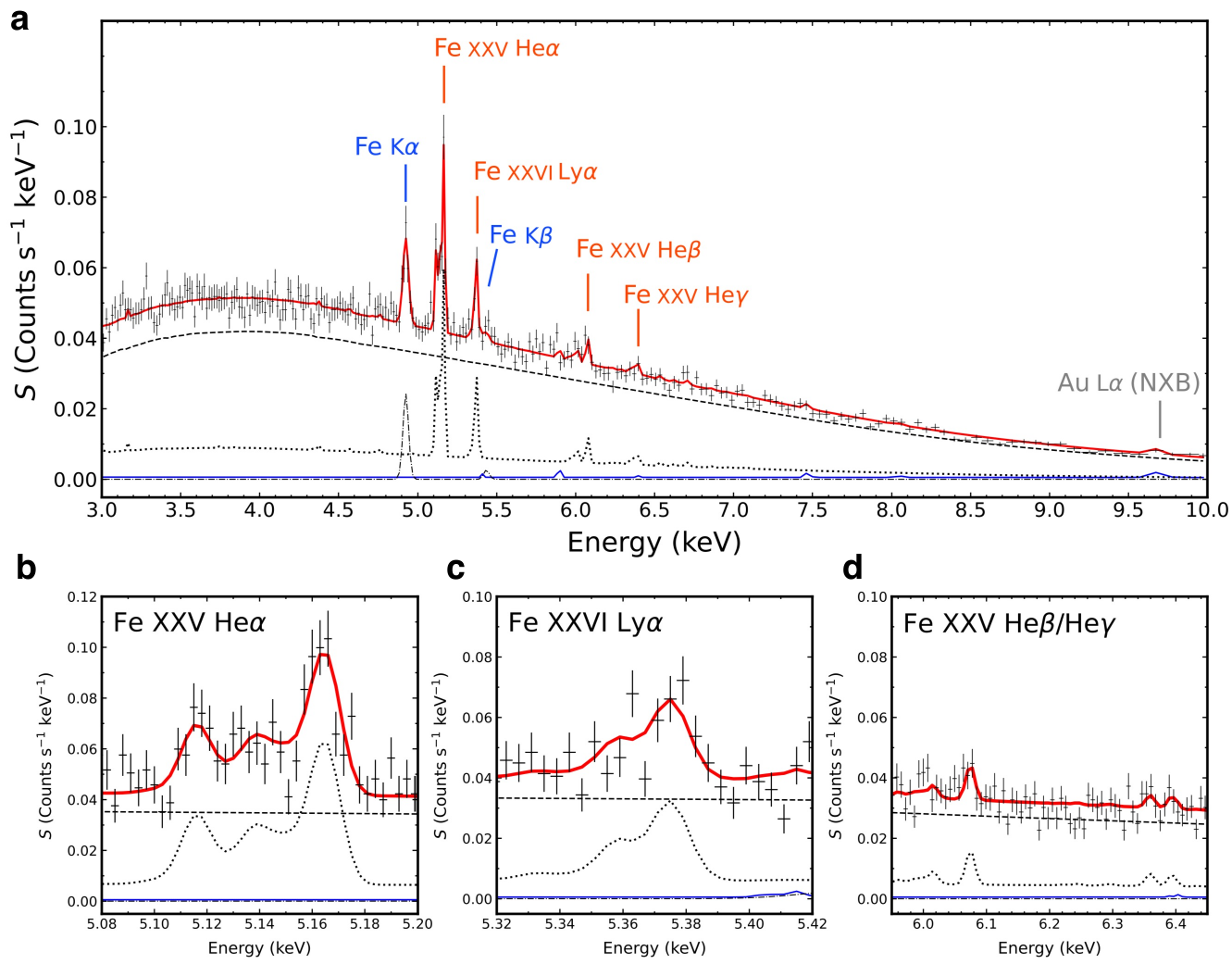}
    \vspace{0.4 cm}
    \caption{
    {\label{Ext_Fig1}\bf Broadband and highly ionized line spectra from the full Resolve field of view.} 
    {\bf a,} The 3--10 keV spectrum with $1\sigma$ uncertainties (black crosses) and the best-fit model (red curve). 
    Blue lines mark the neutral Fe K$\alpha$ and K$\beta$ lines, while orange lines indicate highly ionized lines, including Fe\,\textsc{xxv} He$\alpha$, Fe\,\textsc{xxvi} Ly$\alpha$, Fe\,\textsc{xxv} He$\beta$ and Fe\,\textsc{xxv} He$\gamma$.
    The narrow feature at ${\sim}9.7$~keV arises from the NXB. 
    The dashed, dotted, dash-dotted, and solid blue curves represent the AGN power-law, ICM, Fe K lines and NXB components, respectively. 
    {\bf b--d,} Zoomed-in views of the highly ionized lines: {\bf b,} Fe\,\textsc{xxv} He$\alpha$, {\bf c,} Fe\,\textsc{xxvi} Ly$\alpha_1$/Ly$\alpha_2$ and {\bf d,} Fe\,\textsc{xxv} He$\beta$ and He$\gamma$.
    }
\end{figure*}

\begin{figure*}
    \captionsetup{name=Extended Data Fig.}
    \centering
    \includegraphics[width=0.9\textwidth]{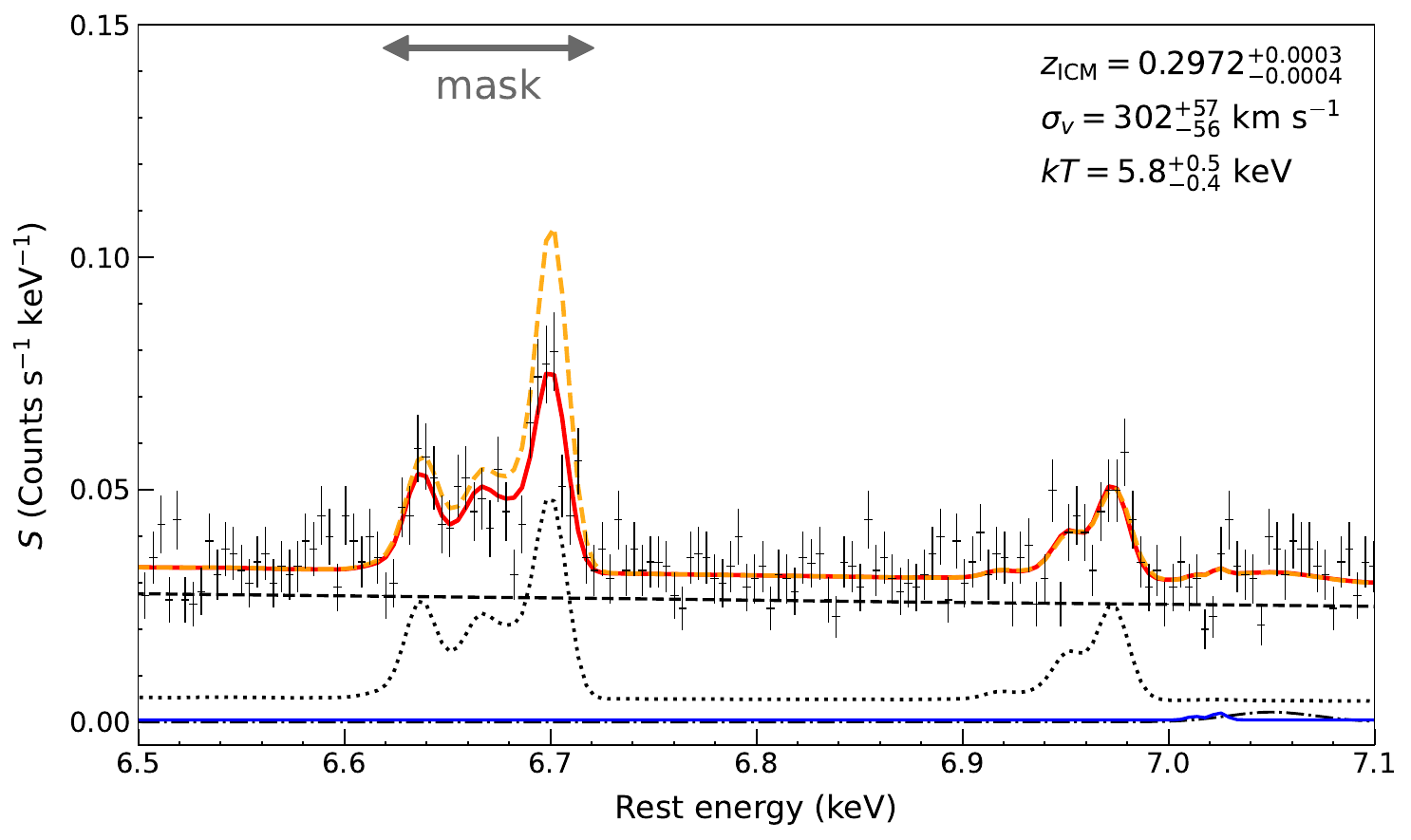}
    \vspace{0.2 cm}
    \caption{
    {\label{Ext_Fig2}\bf Testing the resonance scattering effect.} 
    The rest-frame 6.5--7.1~keV spectrum of Resolve with $1\sigma$ uncertainties (dashed crosses). The orange dashed curve shows the best-fit model when the rest-frame 6.62--6.72~keV band is masked, yielding an ICM redshift of 0.2972, a velocity dispersion of 302~km~s$^{-1}$ and a temperature of 5.8~keV.
    The best-fit model for the full observed-frame 4--7~keV (rest-frame 5.2--9.1~keV) band includes a negative Gaussian to account for the resonance scattering on the w line; the model was adopted as the fiducial model in this study (red). 
    The dashed, dotted, dash-dotted, and solid blue curves represent the AGN power-law, ICM, Fe K lines and NXB components, respectively.
    The corresponding velocity dispersion is $\sigma_v = 283^{+26}_{-29}$~km~s$^{-1}$.
    }    
\end{figure*}

\begin{figure*}[h]
    \captionsetup{name=Extended Data Fig.}
    \centering
    \includegraphics[width=0.85\textwidth]{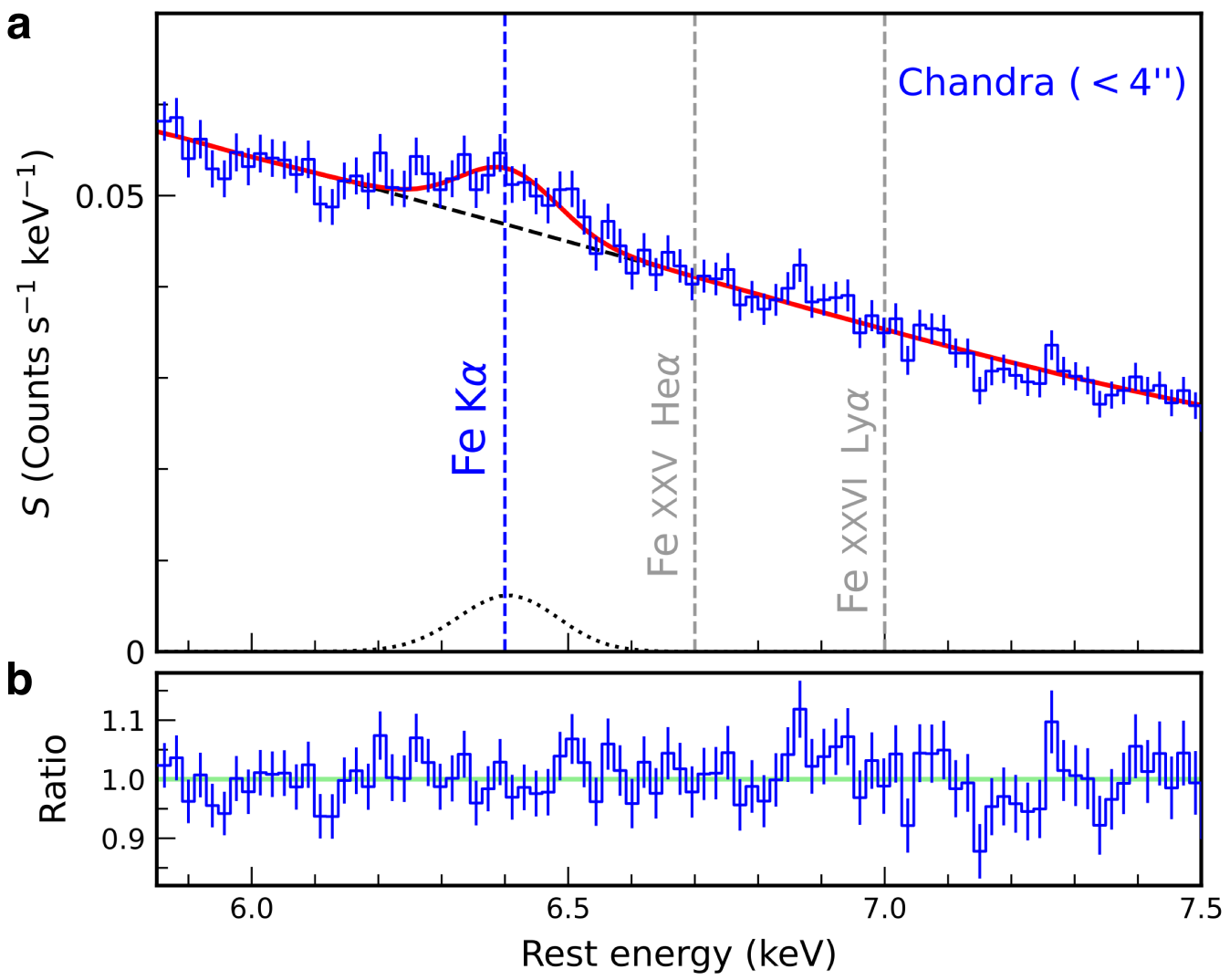}
    \vspace{0.2 cm}
    \caption{
    {\label{Ext_Fig3}\bf Nuclear Chandra spectrum within a radius of 4$''$.} 
    {\bf a,} The rest-frame nuclear Chandra spectrum (blue) and the best-fit model (red) consisting of a power-law continuum (dashed curve) plus a single Gaussian representing the Fe K$\alpha$ line at 6.4~keV (dotted curve).
    The expected energies of Fe K$\alpha$, Fe\,\textsc{xxv} He$\alpha$ and Fe\,\textsc{xxvi} Ly$\alpha$ lines are marked by dashed vertical lines.
    {\bf b,} Ratio of the observed data to the best-fit model, demonstrating the absence of highly ionized Fe lines (Fe\,\textsc{xxv} and Fe\,\textsc{xxvi}) in the nuclear spectrum.
    }
\end{figure*}


\clearpage
\begin{thebibliography}{62}
\expandafter\ifx\csname url\endcsname\relax
  \def\url#1{\burl{#1}}\fi
\expandafter\ifx\csname urlprefix\endcsname\relax\def\urlprefix{URL }\fi
\providecommand{\bibinfo}[2]{#2}
\providecommand{\eprint}[2][]{\url{#2}}
\providecommand{\doi}[1]{\url{https://doi.org/#1}}
\bibcommenthead

\bibitem{Tashiro2025}
\bibinfo{author}{{Tashiro}, M.} \emph{et~al.}
\newblock \bibinfo{title}{{X-Ray Imaging and Spectroscopy Mission}}.
\newblock \emph{\bibinfo{journal}{\pasj}} \textbf{\bibinfo{volume}{77}},
  \bibinfo{pages}{S1--S9} (\bibinfo{year}{2025}).

\bibitem{Takahashi2014}
\bibinfo{author}{{Takahashi}, T.} \emph{et~al.}
\newblock \bibinfo{title}{{The ASTRO-H X-ray astronomy satellite}}.
\newblock In: (eds \bibinfo{editor}{{Takahashi}, T.}, \bibinfo{editor}{{den
  Herder}, J.-W.~A.} \& \bibinfo{editor}{{Bautz}, M.})
  \emph{\bibinfo{booktitle}{Space Telescopes and Instrumentation 2014:
  Ultraviolet to Gamma Ray}}, Vol. \bibinfo{volume}{9144} of
  \emph{\bibinfo{series}{Society of Photo-Optical Instrumentation Engineers
  (SPIE) Conference Series}}, \bibinfo{pages}{914425} (\bibinfo{year}{2014}).

\bibitem{Eckart2025}
\bibinfo{author}{{Eckart}, M.~E.} \emph{et~al.}
\newblock \bibinfo{title}{{Energy gain scale calibration of the XRISM Resolve
  microcalorimeter spectrometer: ground calibration results and on-orbit
  comparison}}.
\newblock \emph{\bibinfo{journal}{J. Astron. Telesc. Instrum. Syst.}} 
  \textbf{\bibinfo{volume}{11}},
  \bibinfo{pages}{042018} (\bibinfo{year}{2025}).

\bibitem{Hitomi2016}
\bibinfo{author}{{Hitomi Collaboration}} \emph{et~al.}
\newblock \bibinfo{title}{{The quiescent intracluster medium in the core of the
  Perseus cluster}}.
\newblock \emph{\bibinfo{journal}{\nat}} \textbf{\bibinfo{volume}{535}},
  \bibinfo{pages}{117--121} (\bibinfo{year}{2016}).

\bibitem{Hitomi2017}
\bibinfo{author}{{Hitomi Collaboration}} \emph{et~al.}
\newblock \bibinfo{title}{{Solar abundance ratios of the iron-peak elements in
  the Perseus cluster}}.
\newblock \emph{\bibinfo{journal}{\nat}} \textbf{\bibinfo{volume}{551}},
  \bibinfo{pages}{478--480} (\bibinfo{year}{2017}).

\bibitem{Bahcall1992}
\bibinfo{author}{{Bahcall}, J.~N.}, \bibinfo{author}{{Jannuzi}, B.~T.},
  \bibinfo{author}{{Schneider}, D.~P.}, \bibinfo{author}{{Hartig}, G.~F.} \&
  \bibinfo{author}{{Green}, R.~F.}
\newblock \bibinfo{title}{{The Ultraviolet Absorption Spectrum of the Quasar
  H1821+643 (z = 0.297)}}.
\newblock \emph{\bibinfo{journal}{\apj}} \textbf{\bibinfo{volume}{397}},
  \bibinfo{pages}{68} (\bibinfo{year}{1992}).

\bibitem{Russell2010}
\bibinfo{author}{{Russell}, H.~R.} \emph{et~al.}
\newblock \bibinfo{title}{{The X-ray luminous cluster underlying the bright
  radio-quiet quasar H1821+643}}.
\newblock \emph{\bibinfo{journal}{\mnras}} \textbf{\bibinfo{volume}{402}},
  \bibinfo{pages}{1561--1579} (\bibinfo{year}{2010}).

\bibitem{Hlavacek-Larrondo2013}
\bibinfo{author}{{Hlavacek-Larrondo}, J.} \emph{et~al.}
\newblock \bibinfo{title}{{The rapid evolution of AGN feedback in brightest
  cluster galaxies: switching from quasar-mode to radio-mode feedback}}.
\newblock \emph{\bibinfo{journal}{\mnras}} \textbf{\bibinfo{volume}{431}},
  \bibinfo{pages}{1638--1658} (\bibinfo{year}{2013}).

\bibitem{XRISM25_Abell2029_I}
\bibinfo{author}{{Xrism Collaboration}} \emph{et~al.}
\newblock \bibinfo{title}{{XRISM Reveals Low Nonthermal Pressure in the Core of
  the Hot, Relaxed Galaxy Cluster A2029}}.
\newblock \emph{\bibinfo{journal}{\apjl}} \textbf{\bibinfo{volume}{982}},
  \bibinfo{pages}{L5} (\bibinfo{year}{2025}).

\bibitem{XRISM25_Abell2029_II}
\bibinfo{author}{{Audard}, M.} \emph{et~al.}
\newblock \bibinfo{title}{{Constraining gas motion and non-thermal pressure
  beyond the core of the Abell~2029 galaxy cluster with XRISM}}.
\newblock \emph{\bibinfo{journal}{\pasj}} \textbf{\bibinfo{volume}{77}},
  \bibinfo{pages}{S242--S253} (\bibinfo{year}{2025}).

\bibitem{Rose2025}
\bibinfo{author}{{Rose}, T.} \emph{et~al.}
\newblock \bibinfo{title}{{XRISM Constrains Atmospheric Motion and Turbulent
  Dissipation in the Archetypal Radio-mode Feedback System Hydra-A}}.
\newblock \emph{\bibinfo{journal}{\apj}} \textbf{\bibinfo{volume}{990}},
  \bibinfo{pages}{42} (\bibinfo{year}{2025}).

\bibitem{XRISM26_Virgo}
\bibinfo{author}{{Xrism Collaboration}} \emph{et~al.}
\newblock \bibinfo{title}{{A XRISM/Resolve View of the Dynamics in the Hot
  Gaseous Atmosphere of M87}}.
\newblock \emph{\bibinfo{journal}{\apj}} \textbf{\bibinfo{volume}{998}},
  \bibinfo{pages}{210} (\bibinfo{year}{2026}).

\bibitem{Fujita2025}
\bibinfo{author}{{Fujita}, Y.}, \bibinfo{author}{{Fukushima}, K.},
  \bibinfo{author}{{Sato}, K.}, \bibinfo{author}{{Fukazawa}, Y.} \&
  \bibinfo{author}{{Kondo}, M.}
\newblock \bibinfo{title}{{XRISM observation of the Ophiuchus galaxy cluster:
  Quiescent velocity structure in the dynamically disturbed core}}.
\newblock \emph{\bibinfo{journal}{\pasj}} \textbf{\bibinfo{volume}{77}},
  \bibinfo{pages}{S270--S275} (\bibinfo{year}{2025}).

\bibitem{XRISM25_Centaurus}
\bibinfo{author}{{XRISM Collaboration}} \emph{et~al.}
\newblock \bibinfo{title}{{The bulk motion of gas in the core of the Centaurus
  galaxy cluster}}.
\newblock \emph{\bibinfo{journal}{\nat}} \textbf{\bibinfo{volume}{638}},
  \bibinfo{pages}{365--369} (\bibinfo{year}{2025}).

\bibitem{Gupta2024}
\bibinfo{author}{{Gupta}, K.~K.} \emph{et~al.}
\newblock \bibinfo{title}{{BASS: XLIII. Optical, UV, and X-ray emission
  properties of unobscured Swift/BAT active galactic nuclei}}.
\newblock \emph{\bibinfo{journal}{\aap}} \textbf{\bibinfo{volume}{691}},
  \bibinfo{pages}{A203} (\bibinfo{year}{2024}).

\bibitem{Blundell2001}
\bibinfo{author}{{Blundell}, K.~M.} \& \bibinfo{author}{{Rawlings}, S.}
\newblock \bibinfo{title}{{The Optically Powerful Quasar E1821+643 is
  Associated with a 300 Kiloparsec-Scale FR I Radio Structure}}.
\newblock \emph{\bibinfo{journal}{\apjl}} \textbf{\bibinfo{volume}{562}},
  \bibinfo{pages}{L5--L8} (\bibinfo{year}{2001}).

\bibitem{Shapovalova2016}
\bibinfo{author}{{Shapovalova}, A.~I.} \emph{et~al.}
\newblock \bibinfo{title}{{First Long-term Optical Spectral Monitoring of a
  Binary Black Hole Candidate E1821+643. I. Variability of Spectral Lines and
  Continuum}}.
\newblock \emph{\bibinfo{journal}{\apjs}} \textbf{\bibinfo{volume}{222}},
  \bibinfo{pages}{25} (\bibinfo{year}{2016}).

\bibitem{Fiore2017}
\bibinfo{author}{{Fiore}, F.} \emph{et~al.}
\newblock \bibinfo{title}{{AGN wind scaling relations and the co-evolution of
  black holes and galaxies}}.
\newblock \emph{\bibinfo{journal}{\aap}} \textbf{\bibinfo{volume}{601}},
  \bibinfo{pages}{A143} (\bibinfo{year}{2017}).

\bibitem{Ruan2015}
\bibinfo{author}{{Ruan}, J.~J.}, \bibinfo{author}{{McQuinn}, M.} \&
  \bibinfo{author}{{Anderson}, S.~F.}
\newblock \bibinfo{title}{{Detection of Quasar Feedback from the Thermal
  Sunyaev-Zel{\textquoteright}dovich Effect in Planck}}.
\newblock \emph{\bibinfo{journal}{\apj}} \textbf{\bibinfo{volume}{802}},
  \bibinfo{pages}{135} (\bibinfo{year}{2015}).

\bibitem{Lacy2019}
\bibinfo{author}{{Lacy}, M.} \emph{et~al.}
\newblock \bibinfo{title}{{Direct detection of quasar feedback via the
  Sunyaev-Zeldovich effect}}.
\newblock \emph{\bibinfo{journal}{\mnras}} \textbf{\bibinfo{volume}{483}},
  \bibinfo{pages}{L22--L27} (\bibinfo{year}{2019}).

\bibitem{Venturi2023}
\bibinfo{author}{{Venturi}, G.} \emph{et~al.}
\newblock \bibinfo{title}{{Complex AGN feedback in the Teacup galaxy. A
  powerful ionised galactic outflow, jet-ISM interaction, and evidence for
  AGN-triggered star formation in a giant bubble}}.
\newblock \emph{\bibinfo{journal}{\aap}} \textbf{\bibinfo{volume}{678}},
  \bibinfo{pages}{A127} (\bibinfo{year}{2023}).

\bibitem{Harrison2024}
\bibinfo{author}{{Harrison}, C.~M.} \& \bibinfo{author}{{Ramos Almeida}, C.}
\newblock \bibinfo{title}{{Observational Tests of Active Galactic Nuclei
  Feedback: An Overview of Approaches and Interpretation}}.
\newblock \emph{\bibinfo{journal}{Galaxies}} \textbf{\bibinfo{volume}{12}},
  \bibinfo{pages}{17} (\bibinfo{year}{2024}).

\bibitem{Gilfanov1987}
\bibinfo{author}{{Gilfanov}, M.~R.}, \bibinfo{author}{{Syunyaev}, R.~A.} \&
  \bibinfo{author}{{Churazov}, E.~M.}
\newblock \bibinfo{title}{{Radial Brightness Profiles of Resonance X-Ray Lines
  in Galaxy Clusters}}.
\newblock \emph{\bibinfo{journal}{Soviet Astron. Lett.}}
  \textbf{\bibinfo{volume}{13}}, \bibinfo{pages}{3} (\bibinfo{year}{1987}).

\bibitem{Hitomi2018_RS}
\bibinfo{author}{{Hitomi Collaboration}} \emph{et~al.}
\newblock \bibinfo{title}{{Measurements of resonant scattering in the Perseus
  Cluster core with Hitomi SXS}}.
\newblock \emph{\bibinfo{journal}{\pasj}} \textbf{\bibinfo{volume}{70}},
  \bibinfo{pages}{10} (\bibinfo{year}{2018}).

\bibitem{Majumder2026}
\bibinfo{author}{{Majumder}, A.} \emph{et~al.}
\newblock \bibinfo{title}{{Spectrally Resolved Gas Kinematics in Cygnus A:
  XRISM Detects AGN Jet-induced Velocity Dispersion in Multitemperature Gas}}.
\newblock \emph{\bibinfo{journal}{\apj}} \textbf{\bibinfo{volume}{998}},
  \bibinfo{pages}{160} (\bibinfo{year}{2026}).

\bibitem{XRISM26_Perseus}
\bibinfo{author}{{The Xrism Collaboration}} \emph{et~al.}
\newblock \bibinfo{title}{{Disentangling multiple gas kinematic drivers in the
  Perseus galaxy cluster}}.
\newblock \emph{\bibinfo{journal}{\nat}} \textbf{\bibinfo{volume}{650}},
  \bibinfo{pages}{309--313} (\bibinfo{year}{2026}).

\bibitem{Hitomi2018_SSM}
\bibinfo{author}{{Hitomi Collaboration}} \emph{et~al.}
\newblock \bibinfo{title}{{Atmospheric gas dynamics in the Perseus cluster
  observed with Hitomi}}.
\newblock \emph{\bibinfo{journal}{\pasj}} \textbf{\bibinfo{volume}{70}},
  \bibinfo{pages}{9} (\bibinfo{year}{2018}).

\bibitem{Russell2024}
\bibinfo{author}{{Russell}, H.~R.} \emph{et~al.}
\newblock \bibinfo{title}{{A cooling flow around the low-redshift quasar
  H1821+643}}.
\newblock \emph{\bibinfo{journal}{\mnras}} \textbf{\bibinfo{volume}{528}},
  \bibinfo{pages}{1863--1878} (\bibinfo{year}{2024}).

\bibitem{Bonafede2014}
\bibinfo{author}{{Bonafede}, A.} \emph{et~al.}
\newblock \bibinfo{title}{{A giant radio halo in the cool core cluster
  CL1821+643.}}
\newblock \emph{\bibinfo{journal}{\mnras}} \textbf{\bibinfo{volume}{444}},
  \bibinfo{pages}{L44--L48} (\bibinfo{year}{2014}).

\bibitem{van_Weeren2024}
\bibinfo{author}{{van Weeren}, R.~J.} \emph{et~al.}
\newblock \bibinfo{title}{{LOFAR high-band antenna observations of the Perseus
  cluster: The discovery of a giant radio halo}}.
\newblock \emph{\bibinfo{journal}{\aap}} \textbf{\bibinfo{volume}{692}},
  \bibinfo{pages}{A12} (\bibinfo{year}{2024}).

\bibitem{Govoni2009}
\bibinfo{author}{{Govoni}, F.} \emph{et~al.}
\newblock \bibinfo{title}{{A search for diffuse radio emission in the relaxed,
  cool-core galaxy clusters A1068, A1413, A1650, A1835, A2029, and Ophiuchus}}.
\newblock \emph{\bibinfo{journal}{\aap}} \textbf{\bibinfo{volume}{499}},
  \bibinfo{pages}{371--383} (\bibinfo{year}{2009}).

\bibitem{Prunier2025c}
\bibinfo{author}{{Prunier}, M.}, \bibinfo{author}{{Ubertosi}, F.},
  \bibinfo{author}{{Hlavacek-Larrondo}, J.} \& \bibinfo{author}{{Pillepich},
  A.}
\newblock \bibinfo{title}{{X-ray shocks in the cool cores of galaxy clusters:
  insights from TNG-Cluster}}.
\newblock \emph{\bibinfo{journal}{\mnras}} \textbf{\bibinfo{volume}{544}},
  \bibinfo{pages}{4188--4207} (\bibinfo{year}{2025}).

\bibitem{McNamara2005}
\bibinfo{author}{{McNamara}, B.~R.} \emph{et~al.}
\newblock \bibinfo{title}{{The heating of gas in a galaxy cluster by X-ray
  cavities and large-scale shock fronts}}.
\newblock \emph{\bibinfo{journal}{\nat}} \textbf{\bibinfo{volume}{433}},
  \bibinfo{pages}{45--47} (\bibinfo{year}{2005}).

\bibitem{Husko2024}
\bibinfo{author}{{Hu{\v{s}}ko}, F.}, \bibinfo{author}{{Lacey}, C.~G.},
  \bibinfo{author}{{Schaye}, J.}, \bibinfo{author}{{Nobels}, F. S.~J.} \&
  \bibinfo{author}{{Schaller}, M.}
\newblock \bibinfo{title}{{Winds versus jets: a comparison between black hole
  feedback modes in simulations of idealized galaxy groups and clusters}}.
\newblock \emph{\bibinfo{journal}{\mnras}} \textbf{\bibinfo{volume}{527}},
  \bibinfo{pages}{5988--6020} (\bibinfo{year}{2024}).

\bibitem{Husko2026}
\bibinfo{author}{{Hu{\v{s}}ko}, F.} \emph{et~al.}
\newblock \bibinfo{title}{{A hybrid active galactic nucleus feedback model with
  spinning black holes, winds and jets}}.
\newblock \emph{\bibinfo{journal}{\mnras}} \textbf{\bibinfo{volume}{547}},
  \bibinfo{pages}{stag324} (\bibinfo{year}{2026}).

\bibitem{Schaye2026}
\bibinfo{author}{{Schaye}, J.} \emph{et~al.}
\newblock \bibinfo{title}{{The COLIBRE project: cosmological hydrodynamical
  simulations of galaxy formation and evolution}}.
\newblock \emph{\bibinfo{journal}{\mnras}} \textbf{\bibinfo{volume}{548}},
  \bibinfo{pages}{stag375} (\bibinfo{year}{2026}).

\bibitem{Prunier2025a}
\bibinfo{author}{{Prunier}, M.}, \bibinfo{author}{{Hlavacek-Larrondo}, J.},
  \bibinfo{author}{{Pillepich}, A.}, \bibinfo{author}{{Lehle}, K.} \&
  \bibinfo{author}{{Nelson}, D.}
\newblock \bibinfo{title}{{X-ray cavities in TNG-Cluster: AGN phenomena in the
  full cosmological context}}.
\newblock \emph{\bibinfo{journal}{\mnras}} \textbf{\bibinfo{volume}{536}},
  \bibinfo{pages}{3200--3219} (\bibinfo{year}{2025}).

\end{thebibliography}

\clearpage

\begin{thebibliography}{30}
\expandafter\ifx\csname url\endcsname\relax
  \def\url#1{\burl{#1}}\fi
\expandafter\ifx\csname urlprefix\endcsname\relax\def\urlprefix{URL }\fi
\providecommand{\bibinfo}[2]{#2}
\providecommand{\eprint}[2][]{\url{#2}}
\providecommand{\doi}[1]{\url{https://doi.org/#1}}
\bibcommenthead


\bibitem[38]{Ishisaki2025}
\bibinfo{author}{Ishisaki, Y.} \emph{et~al.}
\newblock \bibinfo{title}{{Resolve instrument onboard XRISM: design,
  integration, and instrument test results}}.
\newblock \emph{\bibinfo{journal}{Journal of Astronomical Telescopes,
  Instruments, and Systems}} \textbf{\bibinfo{volume}{11}},
  \bibinfo{pages}{042023} (\bibinfo{year}{2025}).
\newblock \urlprefix\url{https://doi.org/10.1117/1.JATIS.11.4.042023}.

\bibitem[39]{Kelley2025}
\bibinfo{author}{{Kelley}, R.~L.} \emph{et~al.}
\newblock \bibinfo{title}{{Resolve instrument onboard the X-Ray Imaging and
  Spectroscopy Mission}}.
\newblock \emph{\bibinfo{journal}{Journal of Astronomical Telescopes,
  Instruments, and Systems}} \textbf{\bibinfo{volume}{11}},
  \bibinfo{pages}{042026} (\bibinfo{year}{2025}).

\bibitem[40]{Kilbourne2018}
\bibinfo{author}{{Kilbourne}, C.~A.} \emph{et~al.}
\newblock \bibinfo{title}{{In-flight calibration of Hitomi Soft X-ray
  Spectrometer. (1) Background}}.
\newblock \emph{\bibinfo{journal}{\pasj}} \textbf{\bibinfo{volume}{70}},
  \bibinfo{pages}{18} (\bibinfo{year}{2018}).

\bibitem[41]{Mochizuki2025}
\bibinfo{author}{{Mochizuki}, Y.} \emph{et~al.}
\newblock \bibinfo{title}{{Optimization of X-ray event screening using ground
  and in-orbit data for the Resolve instrument onboard the XRISM satellite}}.
\newblock \emph{\bibinfo{journal}{Journal of Astronomical Telescopes,
  Instruments, and Systems}} \textbf{\bibinfo{volume}{11}},
  \bibinfo{pages}{042002} (\bibinfo{year}{2025}).

\bibitem[42]{Porter2025}
\bibinfo{author}{{Porter}, F.~S.} \emph{et~al.}
\newblock \bibinfo{title}{{In-flight performance of the XRISM/Resolve detector
  system}}.
\newblock \emph{\bibinfo{journal}{Journal of Astronomical Telescopes,
  Instruments, and Systems}} \textbf{\bibinfo{volume}{11}},
  \bibinfo{pages}{042016} (\bibinfo{year}{2025}).

\bibitem[43]{Ishisaki2018}
\bibinfo{author}{{Ishisaki}, Y.} \emph{et~al.}
\newblock \bibinfo{title}{{In-flight performance of pulse-processing system of
  the ASTRO-H/Hitomi soft x-ray spectrometer}}.
\newblock \emph{\bibinfo{journal}{Journal of Astronomical Telescopes,
  Instruments, and Systems}} \textbf{\bibinfo{volume}{4}},
  \bibinfo{pages}{011217} (\bibinfo{year}{2018}).

\bibitem[44]{Mizumoto2025}
\bibinfo{author}{{Mizumoto}, M.} \emph{et~al.}
\newblock \bibinfo{title}{{High-count-rate effects in event processing for the
  XRISM/Resolve X-ray microcalorimeter. II. Energy scale and resolution in
  orbit}}.
\newblock \emph{\bibinfo{journal}{\pasj}} \textbf{\bibinfo{volume}{77}},
  \bibinfo{pages}{S39--S49} (\bibinfo{year}{2025}).

\bibitem[45]{Hitomi2018_Atomic}
\bibinfo{author}{{Hitomi Collaboration}} \emph{et~al.}
\newblock \bibinfo{title}{{Atomic data and spectral modeling constraints from
  high-resolution X-ray observations of the Perseus cluster with Hitomi}}.
\newblock \emph{\bibinfo{journal}{\pasj}} \textbf{\bibinfo{volume}{70}},
  \bibinfo{pages}{12} (\bibinfo{year}{2018}).

\bibitem[46]{Fruscione2006}
\bibinfo{author}{{Fruscione}, A.} \emph{et~al.}
\newblock \bibinfo{editor}{{Silva}, D.~R.} \& \bibinfo{editor}{{Doxsey}, R.~E.}
  (eds) \emph{\bibinfo{title}{{CIAO: Chandra's data analysis system}}}.
\newblock (eds \bibinfo{editor}{{Silva}, D.~R.} \& \bibinfo{editor}{{Doxsey},
  R.~E.}) \emph{\bibinfo{booktitle}{Observatory Operations: Strategies,
  Processes, and Systems}}, Vol. \bibinfo{volume}{6270} of
  \emph{\bibinfo{series}{Society of Photo-Optical Instrumentation Engineers
  (SPIE) Conference Series}}, \bibinfo{pages}{62701V} (\bibinfo{year}{2006}).

\bibitem[47]{Cash1979}
\bibinfo{author}{{Cash}, W.}
\newblock \bibinfo{title}{{Parameter estimation in astronomy through
  application of the likelihood ratio.}}
\newblock \emph{\bibinfo{journal}{\apj}} \textbf{\bibinfo{volume}{228}},
  \bibinfo{pages}{939--947} (\bibinfo{year}{1979}).

\bibitem[48]{Arnaud1996}
\bibinfo{author}{{Arnaud}, K.~A.}
\newblock \emph{\bibinfo{title}{{XSPEC: The First Ten Years}}}, Vol.
  \bibinfo{volume}{101} of \emph{\bibinfo{series}{Astronomical Society of the
  Pacific Conference Series}}, \bibinfo{pages}{17} (\bibinfo{year}{1996}).

\bibitem[49]{Lodders2009}
\bibinfo{author}{{Lodders}, K.}, \bibinfo{author}{{Palme}, H.} \&
  \bibinfo{author}{{Gail}, H.~P.}
\newblock \bibinfo{title}{{Abundances of the Elements in the Solar System}}.
\newblock \emph{\bibinfo{journal}{Landolt B{\"o}rnstein}}
  \textbf{\bibinfo{volume}{4B}}, \bibinfo{pages}{712} (\bibinfo{year}{2009}).

\bibitem[50]{Akaike1974}
\bibinfo{author}{{Akaike}, H.}
\newblock \bibinfo{title}{{A New Look at the Statistical Model
  Identification}}.
\newblock \emph{\bibinfo{journal}{IEEE Transactions on Automatic Control}}
  \textbf{\bibinfo{volume}{19}}, \bibinfo{pages}{716--723}
  (\bibinfo{year}{1974}).

\bibitem[51]{Birzan2012}
\bibinfo{author}{{B{\^\i}rzan}, L.} \emph{et~al.}
\newblock \bibinfo{title}{{The duty cycle of radio-mode feedback in complete
  samples of clusters}}.
\newblock \emph{\bibinfo{journal}{\mnras}} \textbf{\bibinfo{volume}{427}},
  \bibinfo{pages}{3468--3488} (\bibinfo{year}{2012}).

\bibitem[52]{Russell2013}
\bibinfo{author}{{Russell}, H.~R.} \emph{et~al.}
\newblock \bibinfo{title}{{Radiative efficiency, variability and Bondi
  accretion on to massive black holes: the transition from radio AGN to quasars
  in brightest cluster galaxies}}.
\newblock \emph{\bibinfo{journal}{\mnras}} \textbf{\bibinfo{volume}{432}},
  \bibinfo{pages}{530--553} (\bibinfo{year}{2013}).

\bibitem[53]{Perez-Torres2009}
\bibinfo{author}{{P{\'e}rez-Torres}, M.~A.} \emph{et~al.}
\newblock \bibinfo{title}{{The origin of the diffuse non-thermal X-ray and
  radio emission in the Ophiuchus cluster of galaxies}}.
\newblock \emph{\bibinfo{journal}{\mnras}} \textbf{\bibinfo{volume}{396}},
  \bibinfo{pages}{2237--2248} (\bibinfo{year}{2009}).

\bibitem[54]{Rafferty2006}
\bibinfo{author}{{Rafferty}, D.~A.}, \bibinfo{author}{{McNamara}, B.~R.},
  \bibinfo{author}{{Nulsen}, P.~E.~J.} \& \bibinfo{author}{{Wise}, M.~W.}
\newblock \bibinfo{title}{{The Feedback-regulated Growth of Black Holes and
  Bulges through Gas Accretion and Starbursts in Cluster Central Dominant
  Galaxies}}.
\newblock \emph{\bibinfo{journal}{\apj}} \textbf{\bibinfo{volume}{652}},
  \bibinfo{pages}{216--231} (\bibinfo{year}{2006}).

\bibitem[55]{Nulsen2005}
\bibinfo{author}{{Nulsen}, P.~E.~J.}, \bibinfo{author}{{McNamara}, B.~R.},
  \bibinfo{author}{{Wise}, M.~W.} \& \bibinfo{author}{{David}, L.~P.}
\newblock \bibinfo{title}{{The Cluster-Scale AGN Outburst in Hydra A}}.
\newblock \emph{\bibinfo{journal}{\apj}} \textbf{\bibinfo{volume}{628}},
  \bibinfo{pages}{629--636} (\bibinfo{year}{2005}).

\bibitem[56]{Forman2017}
\bibinfo{author}{{Forman}, W.} \emph{et~al.}
\newblock \bibinfo{title}{{Partitioning the Outburst Energy of a Low Eddington
  Accretion Rate AGN at the Center of an Elliptical Galaxy: The Recent 12 Myr
  History of the Supermassive Black Hole in M87}}.
\newblock \emph{\bibinfo{journal}{\apj}} \textbf{\bibinfo{volume}{844}},
  \bibinfo{pages}{122} (\bibinfo{year}{2017}).

\bibitem[57]{Giacintucci2020}
\bibinfo{author}{{Giacintucci}, S.} \emph{et~al.}
\newblock \bibinfo{title}{{Discovery of a Giant Radio Fossil in the Ophiuchus
  Galaxy Cluster}}.
\newblock \emph{\bibinfo{journal}{\apj}} \textbf{\bibinfo{volume}{891}},
  \bibinfo{pages}{1} (\bibinfo{year}{2020}).

\bibitem[58]{McNamara2011}
\bibinfo{author}{{McNamara}, B.~R.}, \bibinfo{author}{{Rohanizadegan}, M.} \&
  \bibinfo{author}{{Nulsen}, P.~E.~J.}
\newblock \bibinfo{title}{{Are Radio Active Galactic Nuclei Powered by
  Accretion or Black Hole Spin?}}
\newblock \emph{\bibinfo{journal}{\apj}} \textbf{\bibinfo{volume}{727}},
  \bibinfo{pages}{39} (\bibinfo{year}{2011}).

\bibitem[59]{EHT2019}
\bibinfo{author}{{Event Horizon Telescope Collaboration}} \emph{et~al.}
\newblock \bibinfo{title}{{First M87 Event Horizon Telescope Results. I. The
  Shadow of the Supermassive Black Hole}}.
\newblock \emph{\bibinfo{journal}{\apjl}} \textbf{\bibinfo{volume}{875}},
  \bibinfo{pages}{L1} (\bibinfo{year}{2019}).

\bibitem[60]{Arzoumanian2021}
\bibinfo{author}{{Arzoumanian}, Z.} \emph{et~al.}
\newblock \bibinfo{title}{{The NANOGrav 11 yr Data Set: Limits on Supermassive
  Black Hole Binaries in Galaxies within 500 Mpc}}.
\newblock \emph{\bibinfo{journal}{\apj}} \textbf{\bibinfo{volume}{914}},
  \bibinfo{pages}{121} (\bibinfo{year}{2021}).

\bibitem[61]{Vasudevan2007}
\bibinfo{author}{{Vasudevan}, R.~V.} \& \bibinfo{author}{{Fabian}, A.~C.}
\newblock \bibinfo{title}{{Piecing together the X-ray background: bolometric
  corrections for active galactic nuclei}}.
\newblock \emph{\bibinfo{journal}{\mnras}} \textbf{\bibinfo{volume}{381}},
  \bibinfo{pages}{1235--1251} (\bibinfo{year}{2007}).

\bibitem[62]{Duras2020}
\bibinfo{author}{{Duras}, F.} \emph{et~al.}
\newblock \bibinfo{title}{{Universal bolometric corrections for active galactic
  nuclei over seven luminosity decades}}.
\newblock \emph{\bibinfo{journal}{\aap}} \textbf{\bibinfo{volume}{636}},
  \bibinfo{pages}{A73} (\bibinfo{year}{2020}).

\bibitem[63]{Shirotori2026}
\bibinfo{author}{{Shirotori}, M.} \& \bibinfo{author}{{Fujita}, Y.}
\newblock \bibinfo{title}{{XRISM mock observations of simulated active galactic
  nucleus jets in the core of a galaxy cluster}}.
\newblock \emph{\bibinfo{journal}{\pasj}} \textbf{\bibinfo{volume}{78}},
  \bibinfo{pages}{609--616} (\bibinfo{year}{2026}).

\bibitem[64]{Truong2024}
\bibinfo{author}{{Truong}, N.} \emph{et~al.}
\newblock \bibinfo{title}{{X-ray-inferred kinematics of the core intracluster
  medium in Perseus-like clusters: Insights from the TNG-Cluster simulation}}.
\newblock \emph{\bibinfo{journal}{\aap}} \textbf{\bibinfo{volume}{686}},
  \bibinfo{pages}{A200} (\bibinfo{year}{2024}).

\end{thebibliography}

\clearpage
\backmatter
\setcounter{figure}{0}

\section*{Supplementary Information}

\bmhead{Basic properties of H1821+643 and other clusters}
To highlight the distinctive properties of H1821+643 compared to other clusters, we summarised key physical parameters of each cluster and its central AGN in Supplementary Table\,\ref{Sup_Tab1}. These included the 2--10~keV X-ray luminosity of the cool core ($L_{\rm cool}$), its radius ($r_{\rm cool}$) \cite{Russell2024,Birzan2012}, absorption-corrected 2--10~keV AGN luminosity ($L_{\rm AGN,X}$) \cite{Hlavacek-Larrondo2013,Russell2013,Perez-Torres2009,XRISM25_Centaurus}, mechanical energy ($E_{\rm cav}$) and power ($P_{\rm cav}$) of X-ray cavities \cite{Russell2024,Rafferty2006,Nulsen2005,Forman2017,Giacintucci2020}, 1.4~GHz radio luminosity ($L_{\rm 1.4GHz}$) as a tracer of jet activity \cite{Hlavacek-Larrondo2013,Birzan2012} and mass of the central SMBH ($M_{\rm BH}$) \cite{Shapovalova2016,McNamara2011,EHT2019,Arzoumanian2021}.

The 2--10~keV AGN luminosity of H1821+643 exceeded that of other clusters by more than 1--5 orders of magnitude, making it exceptional.  Previous studies have shown that the bolometric-to-X-ray luminosity ratio  ($L_{\rm bol}/L_{\rm AGN,X}$) is typically $\sim$20 and increases with $L_{\rm AGN,X}$ \cite{Vasudevan2007,Duras2020}, implying an even stronger bolometric output in luminous quasars.
The bolometric luminosity of H1821+643, estimated as $50L_{\rm AGN,X}$ \cite{Russell2024}, was consistent with independent multiwavelength 
spectral-energy-distribution fits \citep{Gupta2024}. 
H1821+643 accreted at an Eddington ratio of $\lambda_{\rm Edd} \sim 0.3$--0.6, typical of quasar-mode systems, whereas other clusters typically exhibited $\lambda_{\rm Edd} \sim 0.01$ or lower.
\\

\bmhead{Velocity dispersion and non-thermal energy fraction}
In nearby clusters with less luminous AGNs, XRISM and Hitomi observations showed ICM velocity dispersions of ${\lesssim}160$~km~s$^{-1}$ within the central ${\lesssim}100$~kpc (Perseus \cite{Hitomi2016,XRISM26_Perseus}, Abell~2029 \cite{XRISM25_Abell2029_I,XRISM25_Abell2029_II}, Hydra~A \cite{Rose2025}, Virgo \cite{XRISM26_Virgo}, Ophiuchus \cite{Fujita2025} and Centaurus \cite{XRISM25_Centaurus}). In comparison, Cygnus~A \cite{Majumder2026} and H1821+643 exhibit substantially higher velocity dispersions of ${\sim}$250--300~km~s$^{-1}$ at $kT \sim 6$~keV.

We note that the XRISM observations of the Virgo cluster indicate elevated velocity dispersions only within the innermost few kpc ($<$4.65 kpc) and at low temperatures ($<$2 keV), corresponding to the cold ICM phase \cite{XRISM26_Virgo}. This very narrow region probes physical conditions distinct from the hot ($>$2~keV) ICM on larger scales, and we therefore treat it as an exception not included in our main comparison on $\sim$100~kpc scales.
Measurements at larger radii ($\sim$5--25~kpc) in Virgo are included in our comparison.
Similarly, in Cygnus A, a velocity dispersion of $\sim$440~km~s$^{-1}$ has been reported within the central $<$35~kpc \cite{Majumder2026}, corresponding to the cold ICM ($\lesssim$2~keV) and thus not directly comparable to the hot ICM on $\sim$100~kpc scales. For the hot ICM ($kT \sim 5.5$~keV), a velocity dispersion of $\sim$260~km~s$^{-1}$ has been reported; in this powerful FR II radio galaxy, this is likely associated with bulk motions driven by the expansion of radio cocoons rather than isotropic turbulence \cite{Shirotori2026}.

\begin{figure*}[t]
    \captionsetup{name=Supplementary Fig.}
    \centering
    \includegraphics[width=0.9\textwidth]{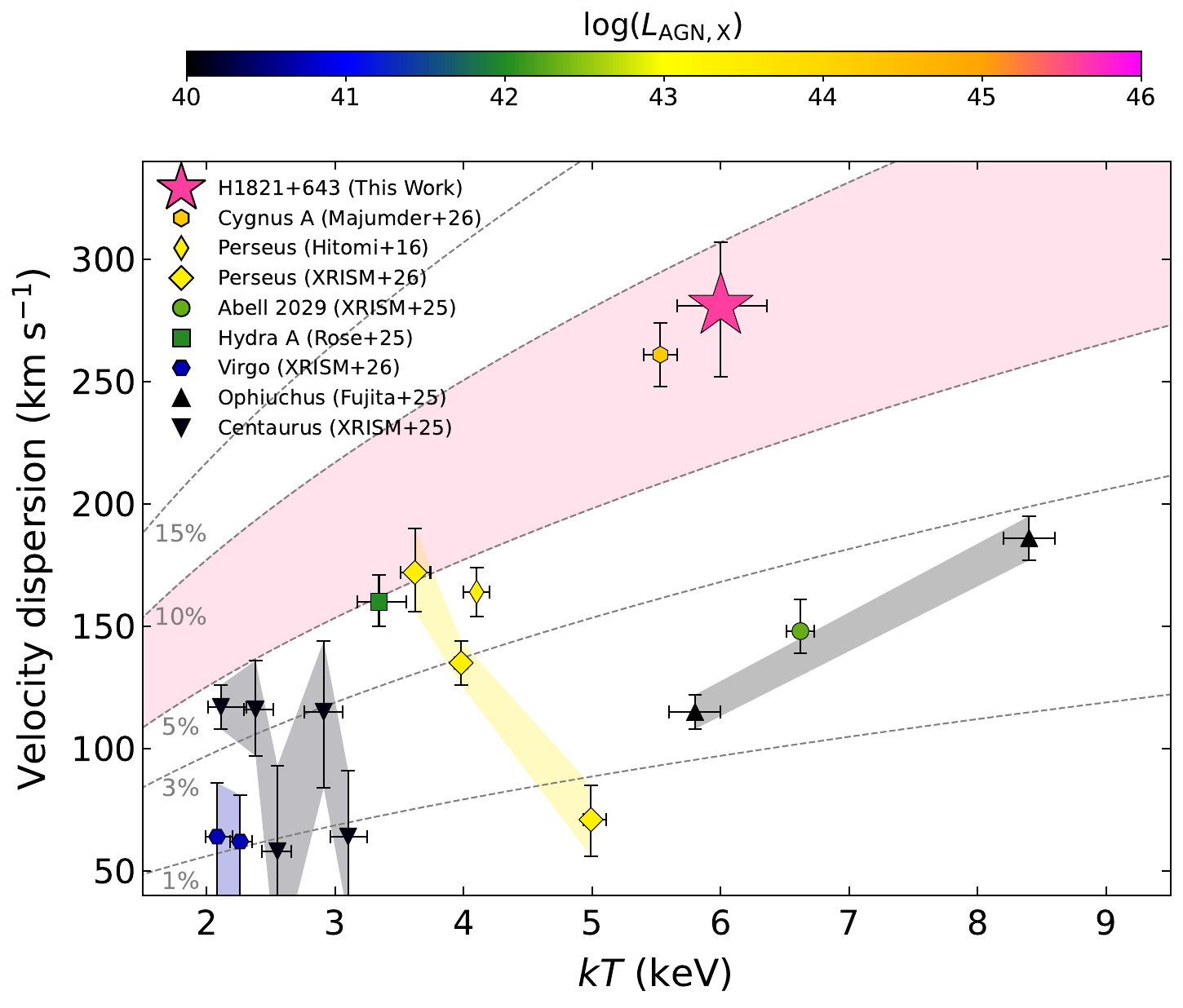}
    \vspace{0.2 cm}
    \caption{
    {\label{Sup_Fig1}\bf Velocity dispersion of the ICM within 100~kpc scales as a function of plasma temperature.} 
    Individual measurements for H1821+643, Cygnus~A, Perseus, Abell~2029, Hydra~A, Virgo, Ophiuchus and Centaurus clusters are shown for the hot ($>$2 keV) ICM. 
    The symbol colour indicates the 2--10~keV AGN luminosity ($L_{\rm X,AGN}$) as shown by the colour bar. 
    For clusters with multiple measurements at different regions, the corresponding ranges are indicated by shaded areas. 
    Error bars represent $1\sigma$ uncertainties.
    Dashed curves indicate combinations of temperature and velocity dispersion corresponding to non-thermal energy fractions $f_{\rm nth}$ of 1\%, 3\%, 5\%, 10\% and 15\%. 
    H1821+643 is the only cluster located in the $f_{\rm nth} \sim 5$--10\% regime (pink shaded region).
    }
\end{figure*}

The non-thermal-to-thermal energy fraction, $f_{\rm nth}$, was estimated from the line-of-sight velocity dispersion by assuming isotropic turbulence or velocity shear, $\sigma_{\rm 3D} = \sqrt{3} \sigma_v$. The resulting non-thermal pressure is then compared with the thermal pressure using $f_{\rm nth} \simeq (1/3) \mu m_p \sigma_{\rm 3D}^2 / kT$, where $\mu \simeq 0.61$ is the mean molecular weight per particle (electrons plus ions) \cite{Truong2024}. This yields $f_{\rm nth} \sim 8.4\%$ in H1821+643, substantially higher than in the other clusters. Whereas clusters with $L_{\rm X,AGN} \lesssim 10^{45}$~erg~s$^{-1}$ show no clear dependence of $f_{\rm nth}$ on AGN X-ray luminosity, with H1821+643 standing out (Fig.\,\ref{Fig3} and Supplementary Fig.\,\ref{Sup_Fig1}).

\begin{figure*}[t]
    \captionsetup{name=Supplementary Fig.}
    \centering
    \includegraphics[width=0.9\textwidth]{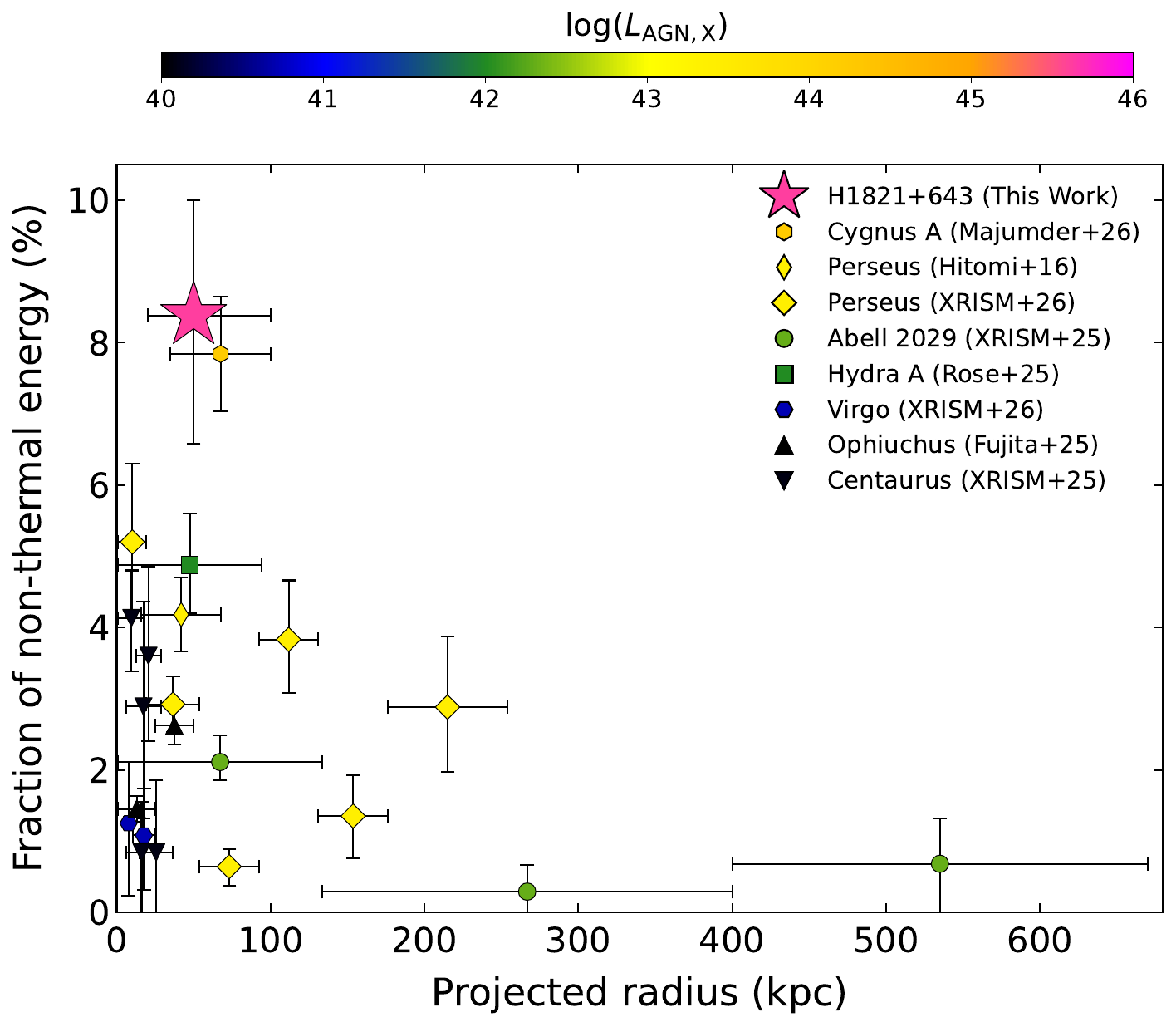}
    \vspace{0.2 cm}
    \caption{
    {\label{Sup_Fig2}\bf Radial distribution of the non-thermal energy fraction in the ICM.}
    The fraction of non-thermal energy, $f_{\rm nth}$, is shown as a function of projected radius.
    Individual measurements for H1821+643, Cygnus~A, Perseus, Abell~2029, Hydra~A, Virgo, Ophiuchus and Centaurus clusters are included for the hot ($>$2~keV) ICM.
    The symbol colour indicates the 2--10~keV AGN luminosity ($L_{\rm X,AGN}$) as shown by the colour bar.
    Different marker types distinguish individual measurements.
    Horizontal bars indicate the projected radial ranges, while vertical error bars represent $1\sigma$ uncertainties.
    }
\end{figure*}

Examining $f_{\rm nth}$ as a function of projected radius, H1821+643 exhibits the most vigorous gas motions among well-studied systems at comparable spatial scales, including Hydra~A (with powerful jets) \cite{Rose2025} and the cool-core cluster Abell~2029 \cite{XRISM25_Abell2029_II}. In other clusters where measurements extend to larger radii (${\gtrsim}100$~kpc), $f_{\rm nth}$ generally decreases. Clusters with lower $L_{\rm X,AGN}$, such as Centaurus \cite{XRISM25_Centaurus} and Ophiuchus \cite{Fujita2025}, are typically measured only within their innermost cores. At similar radii, the data tentatively suggest that $f_{\rm nth}$ increases with AGN luminosity (Supplementary Fig.\,\ref{Sup_Fig2}).
\\

\begin{table*}[h]
\captionsetup{name=Supplementary Table}
{
\caption{\label{Sup_Tab1}\bf Basic properties of the clusters in our sample.}
\centering
\small
\begin{tabular}{lccccccc}
\hline\hline
Cluster name & $L_{\rm cool}$ & $r_{\rm cool}$ & $L_{\rm AGN,X}$ & $E_{\rm cav}$ & $P_{\rm cav}$/$L_{\rm 1.4GHz}$ & $M_{\rm BH}$ \\
 & ($10^{42}$ erg s$^{-1}$) & (kpc) & ($10^{42}$ erg s$^{-1}$) & ($10^{58}$ erg) & ($10^{42}$ erg s$^{-1}$) & ($10^{9} M_{\odot}$) \\
\hline
H1821+643 & ${\approx}3500$ & 90 & $4200^{+100}_{-100}$ & 60 & 700/3.2 & 3.0 \\
Cygnus~A & $293^{+3}_{-3}$ & 92 & $173^{+29}_{-19}$ & 84 & 1300/152.9 & 2.7 \\
Perseus & $545^{+2}_{-2}$ & 110 & $25^{+54}_{-17}$ & 19 & 150/0.24 & 0.34 \\
Abell~2029 & $1049^{+5}_{-5}$ & 127 & $2.1^{+0.2}_{-0.2}$ & 4.8 & 87/0.15 & 4.0 \\
Hydra~A & $227^{+2}_{-2}$ & 107 & $1.0^{+0.1}_{-0.1}$ & 900 & 2000/4.21 & 5.8 \\
Virgo & $6.3^{+0.1}_{-0.1}$ & 39 & $0.051^{+0.003}_{-0.003}$ & 0.5 & 13/0.081 & 6.5 \\
Ophiuchus & $324^{+1}_{-2}$ & 83 & $<$0.33 & 5000 & $\cdots$/0.00072 & 6.9 \\
Centaurus & $22.0^{+0.1}_{-0.1}$ & 87 & ${<}0.012$ & 0.06 & 7.4/0.016 & 2.0 \\
\hline\hline
\end{tabular}
\begin{flushleft}
\textbf{Note.} Key properties of the clusters and their central AGN are listed.
Columns include 
the 2--10~keV luminosity ($L_{\rm cool}$) and 
radius of the cool core ($r_{\rm cool}$) \cite{Russell2024,Birzan2012}, 
intrinsic 2--10~keV AGN luminosity ($L_{\rm AGN,X}$) \cite{Hlavacek-Larrondo2013,Russell2013,Perez-Torres2009,XRISM25_Centaurus}, 
mechanical energy ($E_{\rm cav}$) and 
power of X-ray cavities ($P_{\rm cav}$) \cite{Russell2024,Rafferty2006,Nulsen2005,Forman2017,Giacintucci2020}, 
1.4~GHz radio luminosity ($L_{\rm 1.4GHz}$) \cite{Hlavacek-Larrondo2013,Birzan2012}, and 
central SMBH mass ($M_{\rm BH}$) \cite{Shapovalova2016,McNamara2011,Arzoumanian2021}.\\
\end{flushleft}
}
\end{table*}

\end{document}